\documentclass[preprint,11pt]{elsarticle}
\usepackage[left=2.5cm,right=2.5cm,top=2cm,bottom=3cm]{geometry}
\usepackage{graphicx} 
\usepackage{siunitx}
\usepackage{placeins}
\usepackage{arydshln}
\usepackage{amsmath}

\usepackage{hyperref}
\hypersetup{
    colorlinks=true,
    linkcolor=blue,
    filecolor=magenta,      
    urlcolor=cyan,
    citecolor=blue,
}

\newcommand{\eg}{e.g.\ }

\newcolumntype{C}[1]{>{\centering\let\newline\\\arraybackslash\hspace{0pt}}m{#1}}

\usepackage{natbib}
\setcitestyle{authoryear,open={(},close={)},citesep={;}}

\journal{}

\usepackage{lineno}

\begin{document}

\begin{frontmatter}

\date{}

\title{Mapping global offshore wind wake losses, layout optimisation potential, and climate change effects}

\author[inst1]{Simon C. Warder}
\author[inst1]{Matthew D. Piggott}

\affiliation[inst1]{organization={Department of Earth Science and Engineering},
            addressline={Imperial College London}, 
            city={London},
            postcode={SW7 2BP}, 
            country={UK}}

\begin{abstract}
This study assesses global offshore wind energy resources, wake-induced losses, array layout optimisation potential and climate change impacts. Global offshore ambient potential is first mapped based on reanalysis data. Wake-induced losses are then estimated using an engineering wake model, revealing that locations with low (high) resource typically experience larger (smaller) percentage losses. However, the specific wind speed distribution is found to be important, with narrower distributions generally leading to greater losses. This is due to the overlap between the wind speed distribution and the high-sensitivity region of the turbine thrust and power curves. Broadly, this leads to much stronger wake-induced losses in the tropics (which experience the trade winds) than mid-latitudes. However, the tropics also experience a narrower wind direction distribution; the results of this study demonstrate that this leads to greater potential for mitigation of wake effects via layout optimisation. Finally, projected changes in wind potential and wake losses due to climate change under a high-emission scenario are assessed. Many regions are projected to decrease in ambient wind resources, and furthermore these regions will typically experience greater wake-induced losses, exacerbating the climate change impact. These results highlight the different challenges and opportunities associated with exploiting offshore wind resources across the globe.
\end{abstract}


\end{frontmatter}


\section{Introduction}

Renewable energy is rapidly expanding across the world \citep{hassan2024comprehensive}. Offshore wind in particular is undergoing a phase of unprecedented growth and falling costs \citep{li2022review}. As of December 2023 there was a total of 75.2 \si{\giga\watt} of offshore wind installed capacity, with a further 138 \si{\giga\watt} expected to be added by 2028 \citep{gwec_2024}, including within a number of emerging markets.

To date, 45\% of global offshore wind has been installed in Europe \citep{gwec_2024}, and a large number of studies have investigated current \citep{decastro2019overview,hasager2020europe,martinez2022mapping} and future \citep{decastro2019overview,martinez2021wind,hahmann2022current,martinez2023evolution} wind resources in this region, as well as implications for energy/power system modelling \citep{fernandez2019offshore,lyden2024pypsa,glaum2024offshore}. Due to the imminent global expansion of offshore wind, there are a growing number of studies considering current and/or future wind resource in other regions such as the USA \citep[e.g.][]{costoya2020suitability}, South America \citep[e.g.][]{pimenta2019brazil,vinhoza2021brazil,shadman2023review}, Asia \citep[e.g.][]{dinh2022offshore,patel2022revised,abdullah2023approach} or globally \citep[e.g.][]{bandoc2018spatial,weiss2018marine,soares2020global}.

A number of popular tools exist that are dedicated to mapping wind resources. These include so-called `wind atlas' products, such as the New European Wind Atlas \citep{hahmann2020making,dorenkamper2020making} and Global Wind Atlas \citep{davis2023global}. Both are based on ERA-5 hindcast data \citep{hersbach2020era5}, dynamically downscaled using the Weather Research and Forecasting (WRF) model \citep{powers2017weather}, with `microscale' modelling via the WAsP software suite \citep{mortensen2001wind}. Reanalysis products are also used directly for wind resource mapping \citep[\eg][]{soares2020global,gruber2022towards}. Either reanalyses or wind atlases are commonly used as inputs into energy systems models (ESMs), which are in turn used to inform policy decisions regarding future energy mix and infrastructure. One popular tool providing ESM wind power inputs is the Renewables.Ninja project \citep{staffell2016using}, which is based on bias-corrected reanalysis data. In the climate change context, wind resource mapping is commonly undertaken using either global projection model outputs such as from CMIP5 \citep[\eg][]{reyers2016future,zheng2019projection} or CMIP6 \citep[\eg][]{ibarra2023cmip6}, or using downscaled climate products such as from the CORDEX project \citep[\eg][]{li2020historical,molina2022added,akperov2023future}, or other downscaling methods \citep[\eg][]{zhang2021future,fernandez2023dynamic}.

An important aspect to assessing potential wind resource extraction is power loss due to turbine wake effects \citep{jensen1983note}. The extraction of kinetic energy by a wind turbine reduces the wind speed immediately downstream, impacting the power output from any downstream turbines within the same farm (so-called `intra-farm' or `internal' wakes). This effect also arises at larger spatial scales, where an upstream wind farm can reduce the power output of a downstream farm \citep{lundquist2019costs,schneemann2020cluster,canadillas2022offshore}, resulting in so-called `inter-farm' or `cluster' wakes. In the intra-farm case, average wake effects will depend on the turbine type and the layout of the wind farm, and in the inter-farm case on the size, density, capacity and relative positions of the farms \citep{wang2023inter}. However, in both cases, another important factor is the local distribution of wind speed and direction. The wind speed is important since wake effects can only impact power output while the wind speed is greater than the turbine's cut-in speed, but not significantly greater than the rated speed. The wind direction is important because a given pair of turbines or farms only interact when the wind direction is (broadly) aligned with their relative positions.

Intra-farm wake effects are well studied in the context of optimal layout design \citep{elkinton2008algorithms,samorani2013wind,shakoor2016wake,PiggottReview2022} and optimal wind farm control \citep{menezes2018review,kheirabadi2019quantitative}. A number of modelling and optimisation approaches have emerged for the intra-farm wake problem \citep{goccmen2016wind}, including via CFD simulations \citep{sanderse2011review}, so-called `engineering' wake models \citep{archer2018review}, or more recently using machine learning surrogate models trained using either CFD model-generated training data \citep[e.g.][]{ti2020wake,zhang2020novel,purohit2022evaluation,li2023end}, or directly from observation data \citep[e.g.][]{ashwin2022data}.

However, despite the breadth of literature on wake effects, their incorporation into resource assessment tools or studies is limited. In the climate change context, \citet{hahmann2022current} use an engineering wake model to include wake-induced losses of a proposed wind farm cluster, within a study of the climate impact on wind energy resources in the North Sea. \citet{warder2024climate} study the evolution of wake effects (on various spatial scales) with climate change, based on a near-future set of wind farms in the German Bight, finding statistically significant changes in wake effects during the summer months. In the reanalysis context, and in particular for models used as inputs into ESMs, wake effects may be implicitly accounted for via bias correction methods based on matching models to observed capacity factors \citep{staffell2016using,benmoufok4856203improving}. However, in this case, variation in wake effects either spatially or due to differing wind farm designs would be neglected. Alternatively, wake effects are sometimes explicitly accounted for, \eg via a fixed `array efficiency' factor \citep{bosch2018temporally,bosch2019global}, or using pre-computed losses for idealised farms of various sizes \citep{glaum2024offshore}, such as those found in \citet{volker2017prospects,energiewende2020making,gea2022value}. In reality, however, wake effects vary depending on the specific wind climate experienced at a given location, as well as with climate change \citep{devis2018should}. Furthermore, the potential to mitigate these losses via array layout optimisation will vary with the wind direction distribution, with narrower distributions likely to offer greater optimisation potential. These factors will have implications for the transfer of knowledge and experience from regions such as Europe, where the offshore wind industry is relatively mature, to emerging markets elsewhere in the world, where the challenges and opportunities associated with wake effects may differ significantly.

Two key gaps are identified in the above literature, which are directly addressed in the present study. Firstly, to the authors' knowledge, there has been no globally consistent assessment of the impact of wake effects on the ability to extract wind power, and the scope to mitigate these losses via array layout optimisation; these factors are essential to the transfer of knowledge to emerging offshore wind markets. Secondly, there has been no global assessment of projected changes in wake-induced losses due to climate change; this has significant implications for long-term planning and policy makers. Specifically, with a focus on global spatial variation, the objectives of this study are to investigate
\begin{enumerate}[(i)]
    \item ambient offshore wind potential;
    \item the power losses due to intra-farm wake effects for a selected wind farm;
    \item the layout optimisation potential, defined as the percentage improvement in wind farm power which can be achieved via layout optimisation, compared with a simple grid layout;
    \item the projected impact of climate change on both ambient offshore wind potential, and intra-farm wake effects, under the SSP585 scenario.
\end{enumerate}

The remainder of the paper is structured as follows. Section \ref{sec:methods} presents the data and methods used within this study, including reanalysis data, climate projection data, and wake modelling and layout optimisation approaches. Corresponding results are presented and discussed in section \ref{sec:results}. Finally, conclusions are drawn in section \ref{sec:conclusions}.

\section{Methods}
\label{sec:methods}

\begin{figure}
    \centering
    \includegraphics[width=0.5\linewidth]{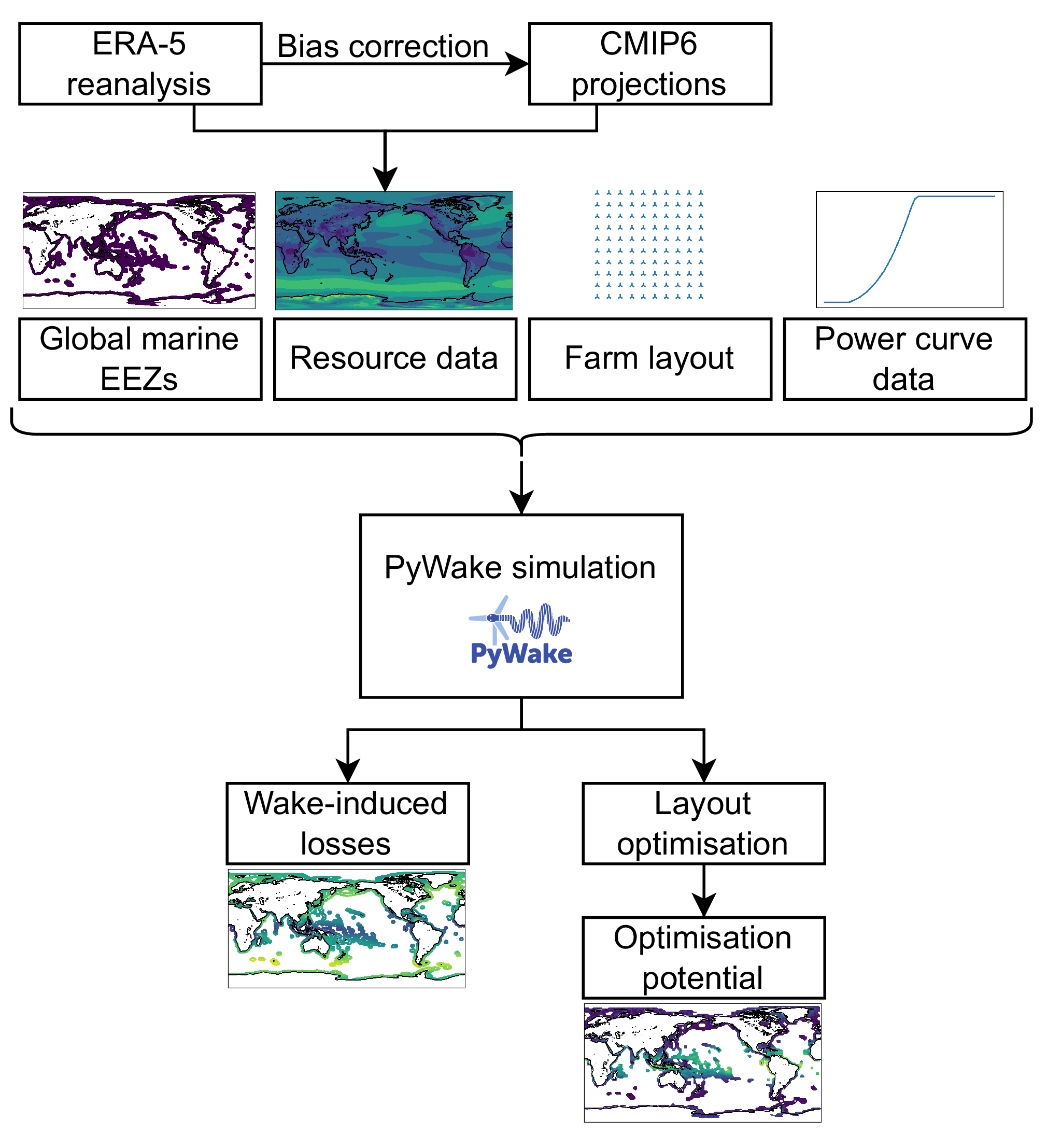}
    \caption{Schematic diagram of the workflow used within this study.}
    \label{fig:workflow_schematic}
\end{figure}

The workflow employed within this study is summarised in figure \ref{fig:workflow_schematic}. Resource data is taken from ERA-5 reanalysis, as well as CMIP6 climate projections. In the latter case, bias correction is performed using ERA-5 data as ground truth. Along with data describing global marine exclusive economic zones (EEZs), and an assumed wind farm layout and turbine power curve, these resource datasets are used to generate global maps of both ambient wind potential, and wake-induced losses. The wake modelling utilises an engineering wake model within the PyWake software suite. Wind turbine layout optimisation is subsequently performed using the TopFarm package, in order to map the `optimisation potential' of wind farms across marine EEZs. Further detail on each of the components of this workflow is given in the following subsections.

\subsection{Data and pre-processing}

\subsubsection{ERA-5}

This study utilises reanalysis data from ERA-5 \citep{hersbach2020era5}, for the historical period 1995--2014. The wind speed and direction are derived from the 100 \si{\metre} $u$ and $v$ wind components from this ERA-5 data. The data has a spatial resolution of 0.25\si{\degree}, and hourly temporal resolution. The data is downloaded from the Copernicus Climate Data Store at a 6-hourly resolution. From this, the wind rose is derived for each grid cell, based on wind speed bins from 0 to 25 \si{\metre\per\second} in steps of 0.5 \si{\metre\per\second}, and directional bins of 10\si{\degree}. The use of a wind rose to represent the wind distribution at each location reduces the total volume of data required to represent global wind climate at high resolution.

\subsubsection{CMIP6}
\label{sec:data_cmip6}

The future climate scenario assessed in this work is derived from CMIP6 projections, and specifically the SSP585 scenario, for the period 2081--2100. This high-emissions scenario corresponds to intensified fossil fuel exploitation, leading to radiative forcing of 8.5 \si{\watt\per\metre\squared} by the year 2100, and lies at the upper end of the scenarios considered in the literature. This scenario is chosen on the basis that it will produce responses of the greatest magnitude, making it easiest to identify robust trends. This future scenario is compared against corresponding CMIP6 data from the `historical' scenario, using the period 1995--2014.

The specific CMIP6 model outputs selected for this study are based on the availability of both the `historical' and SSP585 scenarios, a nominal resolution of 100 \si{\kilo\metre} or finer, and surface wind speeds with daily frequency. While higher spatial and temporal resolution is available for some models, these selection criteria were chosen to ensure that a sufficient range of models was available from which to form an ensemble. Data was searched and downloaded via the Earth System Grid Federation (ESGF) search tool (\url{https://esgf-ui.ceda.ac.uk/cog/search/cmip6-ceda/}). The resulting 10 models are detailed in table \ref{tab:cmip6_models}.

Each of these models provides the `surface' wind speed, at a nominal height of 10 \si{\metre}. Since power estimation requires wind speeds at hub heights, these 10 \si{\metre} data are extrapolated to 100 \si{\metre}, using a power law with an exponent of 0.06, following \citet{carvalho2021wind}, who chose this value based on typical vertical profiles from ERA-5 data, which is available at both 10 and 100 \si{\metre}. The resulting data is converted to a wind rose map (consistent with the treatment of the ERA-5 data described above), then regridded by bilinear interpolation to match the ERA-5 grid.

The climate projection data is further bias-corrected against the ERA-5 data. This bias correction is performed separately for each grid cell. \citet{warder2024climate} compared several common methods for bias correction, with a quantile mapping method based on fitted Weibull distributions found to perform best. However, since wind speeds are not necessarily well described by Weibull distributions at all locations globally, this study employs an empirical quantile mapping method, which was also found by \citet{warder2024climate} to perform well. The use of a quantile mapping method, rather than a simpler rescaling or offset approach, is also well suited here since the climate datasets may underestimate wind speed variability due to the use of daily mean data. A quantile mapping approach is sufficient to address this issue, since it can amend the shape of the wind speed probability distribution.

The bias correction is derived based on the climate dataset over the period that coincides with the historical ERA-5 data, using the latter as `ground truth'. Denoting the empirical cumulative density function (CDF) of the ERA-5 data at a particular grid cell and for a particular wind direction as $F_\text{ERA-5}$, and the corresponding CDF from the `historical' climate dataset as $F_\text{hist}$, the corrected wind speed $w_\text{corr}$ is given as a function of the raw wind speed $w_\text{raw}$ as
\begin{equation}
    \label{eq:eqm_1}
    w_\text{corr} = F^{-1}_\text{ERA-5} \left( F_\text{hist}(w_\text{raw}) \right).
\end{equation}
Since the approach of this study is based on wind roses, the projected climate data is represented as a discrete probability density function (PDF) at a particular set of speeds. Denoting the raw PDF as $f_\text{raw}(w)$ and the corrected PDF as $f_\text{corr}(w)$, we assume that
\begin{equation}
    \label{eq:eqm_2}
    f_\text{corr}(w_\text{corr}) = f_\text{raw}(w_\text{raw}).
\end{equation}
The raw PDF therefore directly provides the corrected PDF, but evaluated at the `corrected' speeds given by equation \eqref{eq:eqm_1}. Linear interpolation is then used to evaluate the corrected PDF on the original set of wind speeds.

\begin{table}[]
    \centering
    \footnotesize
    \begin{tabular}{c|p{5.5cm}|c|c}
        Model name & Institute & Resolution & Citation \\ \hline
        CESM2-WACCM & National Center for Atmospheric Research (NCAR) -- USA & 1.25\si{\degree} lon $\times$ 0.938\si{\degree} lat & \citet{danabasoglu2020community} \\
        CMCC-CM2-SR5 & Euro‐Mediterranean Center on Climate Change (CMCC) Foundation -- Italy & 1.25\si{\degree} lon $\times$ 0.938\si{\degree} lat & \citet{cherchi2019global} \\
        CMCC-ESM2 & Euro‐Mediterranean Center on Climate Change (CMCC) Foundation -- Italy & 1.25\si{\degree} lon $\times$ 0.938\si{\degree} lat & \citet{lovato2022cmip6} \\
        MPI-ESM1-2-HR & Max Planck Institute for Meteorology (MPI-M) -- Germany & 0.938\si{\degree} lat $\times$ 0.938\si{\degree} lon & \citet{muller2018higher} \\
        MRI-ESM2-0 & Meteorological Research Institute (MRI) -- Japan & 1.125\si{\degree} lat $\times$ 1.125\si{\degree} lon & \citet{kawai2019significant} \\
        GFDL-CM4 & Geophysical Fluid Dynamics Laboratory (NOAA-GFDL) -- USA & 1\si{\degree} lat $\times$ 1.25\si{\degree} lon & \citet{held2019structure} \\
        GFDL-ESM4 & Geophysical Fluid Dynamics Laboratory (NOAA-GFDL) -- USA & 1\si{\degree} lat $\times$ 1.25\si{\degree} lon & \citet{dunne2020gfdl} \\
        INM-CM4-8 & Institute for Numerical Mathematics (INM) -- Russia & 1.5\si{\degree} lat $\times$ 2\si{\degree} lon & \citet{volodin2019inm} \\
        INM-CM5-0 & Institute for Numerical Mathematics (INM) -- Russia & 1.5\si{\degree} lat $\times$ 2\si{\degree} lon & \citet{volodin2017simulation} \\
        TaiESM1 & Research Center for Environmental Changes (RCEC) -- Taiwan & 1.25\si{\degree} lat $\times$ 0.938\si{\degree} lon & \citet{lee2020rcec} \\
    \end{tabular}
    \caption{Summary of CMIP6 models used within this study.}
    \label{tab:cmip6_models}
\end{table}

\subsubsection{Marine exclusive economic zones}

This study uses global data for marine exclusive economic zones (EEZs) from \citet{eez_data}, in order to restrict the analysis to plausible offshore regions, and to aggregate results for different countries. However, the inclusion of other practical constraints to wind farm deployment, such as bathymetry, shipping lanes etc, is left to future work. This study uses a resolution of 0.25\si{\degree}, which would not be sufficient to adequately resolve all such features.

\subsection{Mapping wake effects}
\label{sec:methods_mapping_wake_effects}

This study seeks to investigate the influence of wind speed distribution on the magnitude of intra-farm wake effects. The approach is based on computing wind roses at each grid cell whose centroid is within the EEZs. For each such grid cell, two quantities are computed:
\begin{enumerate}[(i)]
    \item the `ambient' capacity factor, or ambient potential, which would be produced by a single turbine within that cell, or equivalently the capacity factor of a wind farm, neglecting all wake effects;
    \item the `waked' capacity factor which would be produced by a `standard' wind farm within that cell, accounting for wake effects.
\end{enumerate}

The reference wind turbine used within this work is the 10 \si{\mega\watt} reference turbine from IEA Task 37 \citep{bortolotti2019iea}, for which power and thrust curve data is available. Neglecting conversion efficiency, the rated (mechanical) power of this turbine is 10.6 \si{\mega\watt}. The turbine has a hub height of 119 \si{\metre} and a rotor diameter of 198 \si{\metre}. As a `standard' farm, this study takes a 10 by 10 square array of such wind turbines, with a spacing of 5 rotor diameters. This farm has a total installed capacity of 1.06 \si{\giga\watt}. The choice of wind turbine model, turbine separation and total capacity were selected in order to be broadly representative of current and near-future offshore wind farms.

The wind rose at a given grid cell is a discrete probability density function, represented by a two-dimensional array which can be denoted $f_{s,d}$, where $s$ and $d$ denote a particular wind speed and direction, respectively. The first quantity above, the ambient capacity factor, does not depend on the wind direction. We therefore introduce the marginal wind speed distribution, given by
\begin{equation}
    f_s = \sum_d f_{s,d}.
\end{equation}
The `ambient' capacity factor at a given location can then be estimated directly from the power curve $P(s)$, as
\begin{equation}
    \text{CF}_\text{ambient} = 100\% \cdot \frac{1}{P_\text{max}} \sum_s f_s P(s_\text{hh}),
\end{equation}
where $P_\text{max}$ is the nominal or rated power of the turbine, in this case 10.6 \si{\mega\watt}, and $s_\text{hh}$ is the wind speed extrapolated to the turbine hub height, which is again performed assuming a power law profile with an exponent of 0.06.

For the second quantity above, the waked capacity factor, the initial focus of this study is on the effect of the wind speed distribution. This can be achieved by taking the marginal wind speed distribution, and for each wind speed, estimating the power produced by the farm, averaged over all possible rotations of the farm. This is equivalent to assuming that all wind directions are equally likely at a given location. This also has the effect of reducing the sensitivity of the waked power with respect to the specific turbine array layout chosen, and the local prevailing wind. I.e., the aim is to exclude variations in wake-induced losses which arise from the alignment (or otherwise) of the principal directions of the farm's grid layout with the prevailing wind.

Estimation of the power output of the wind farm, at a particular 100 \si{\metre} wind speed $s$ and subject to a given rotation (or equivalently a given wind direction) $d$, is simulated using the engineering wake model of \citet{bastankhah2014new}, as implemented within the PyWake software suite \citep{mads_m_pedersen_2019_2562662}. Specifically, the predefined `IEA37SimpleBastankhahGaussian' wind farm model available within PyWake is used. A fixed wake expansion parameter of $k=0.032$ is assumed, neglecting the influence of turbulence (or any other factors) on wake expansion. The influence of air density on power production or wakes is neglected. These assumptions enable the isolation of the influence of wind speed distributions on power and wakes, and further exploration of the impact of air density is left to future work. Turbine interactions are computed by starting with the most upwind turbine and iterating in the downstream direction. Squared sum superposition is used to estimate the effect of overlapping wakes. For further detail regarding the implementation of the wake model, the reader is referred to the PyWake documentation \citep{pywake_docs}. This modelling workflow constitutes a commonly-used approach, and the wake model of \citet{bastankhah2014new} has been previously validated against a variety of observation data including LES simulations \citep{bastankhah2014new}, lidar \citep{carbajo2018wind} and operational data \citep{rodrigo2020validation}.

The simulated power output of our `standard' farm for speed $s$ and rotation (or wind direction) $d$ is denoted $P_{s,d}$. The mean power for each wind speed can then be computed as
\begin{equation}
    P_s = \frac{1}{n_\text{dir}} \sum_d P_{s,d},
\end{equation}
where $n_\text{dir}$ is the number of wind directions, chosen as 36, or equivalently using bins of 10\si{\degree}. This mean power is then combined with the marginal wind speed distribution $f_s$, to produce the final `waked' capacity factor, given by
\begin{equation}
    \label{eq:cf_waked}
    \text{CF}_\text{waked} = 100\% \cdot \frac{1}{N P_\text{max}} \sum_s f_s P_s,
\end{equation}
where $N=100$ is the total number of wind turbines. It should be emphasised that by computing the capacity factor in this way, all information about the wind direction distribution is discarded, isolating the influence of the wind speed distribution. This allows for greater insight into the influence of wind speed distribution on power and wake effects, and reduces the sensitivity of the results to the choice of wind turbine layout within the `standard' farm.

As well as the raw $\text{CF}_\text{ambient}$ and $\text{CF}_\text{waked}$ values, the capacity factor lost to wakes is also presented, given by
\begin{equation}
    \label{eq:cf_lost}
    \text{CF lost} = \text{CF}_\text{waked} - \text{CF}_\text{bg},
\end{equation}
and the percentage of power lost to wakes, given by
\begin{equation}
    \label{eq:pc_power_lost}
    \text{\% losses} = 100\% \cdot (\text{CF}_\text{waked} - \text{CF}_\text{bg}) / \text{CF}_\text{bg}.
\end{equation}

\subsection{Mapping layout optimisation potential}
\label{sec:methods_layout_opt_pot}

The distribution of wind directions at a given location influences the optimal array layout. A highly focused wind direction distribution presents the possibility of mitigating wake losses by placing downstream turbines outside of the wake of upstream turbines. For uniformly distributed wind directions, however, significantly less potential for optimisation is expected.

To explore this, the variation in wind direction spread across EEZs is first investigated, based on the ERA-5 reanalysis data. The circular standard deviation of the wind direction is used as a measure of wind direction spread.

The optimisation potential at each grid cell within the model domain is then estimated by taking the `standard' farm as a starting point, and optimising using the TopFarm Python package, version 2.3.2 \citep{rethore2014topfarm}. This uses PyWake to estimate power for a given array layout, and interfaces with SciPy to perform a gradient-based optimisation of the position of each turbine, to maximise total power output. The optimisation potential is then defined as
\begin{equation}
    \label{eq:opt_potential}
    \text{Opt potential} = 100\% \cdot \frac{\text{CF}_\text{opt} - \text{CF}_\text{waked}}{\text{CF}_\text{waked}},
\end{equation}
where $\text{CF}_\text{waked}$ is given by equation \eqref{eq:cf_waked} and $\text{CF}_\text{opt}$ is obtained by optimisation using TopFarm. Note that for consistency with the other results presented, $\text{CF}_\text{waked}$ is computed by averaging the wake effects over all rotations of the initial wind farm, which has a square grid layout. However, the optimal layout design is based on optimising from a single initial condition in which the farm grid layout is aligned with the mean wind direction.

Since there are approximately 300,000 grid cells within the global EEZs at the resolution considered, the computational cost associated with optimising the array layout of the 100-turbine farm for every grid cell is prohibitive. To mitigate this, the grid is coarsened by a factor of 8 in each dimension, for a total of 4,632 grid cells. The coarsening discards data from unselected grid cells, since spatial averaging would produce artificially smooth wind roses. Although the spatial resolution is decreased, this level of coarsening maintains the ability to observe global trends in optimisation potential, at a manageable computational cost.

\subsection{Mapping future changes due to climate}

The climate projection datasets used were described in section \ref{sec:data_cmip6}, including extrapolation to 100 \si{\metre}, and bias correction against ERA-5 data.

The analysis of climate change effects in this study is restricted to the influence of wind speed distribution. This neglects any effects due to changing wind direction distribution, and also neglects any potential climatic feedback from large-scale wind deployment \citep{wang2023climatic}; although both of these aspects warrant further research, they lie beyond the scope of this study. For each climate model, the method of section \ref{sec:methods_mapping_wake_effects} is used to compute the ambient and `waked' farm capacity factors, for both the `historical' and future periods. From these, the projected change in ambient potential is calculated as
\begin{equation}
    100\% \cdot \frac{\widehat{CF}_\text{ambient} - CF_\text{ambient}}{CF_\text{ambient}},
\end{equation}
where $\widehat{CF}_\text{ambient}$ denotes the ambient CF for the future period and $CF_\text{ambient}$ denotes the ambient CF from the historical period. The projected change in wake-induced losses is then calculated as
\begin{equation}
    \widehat{\text{\% losses}} - \text{\% losses},
\end{equation}
where the hat again indicates the future period.

The results presented within this study are based on the mean changes in the above quantities, across the ensemble of 10 climate model projections. Note that the `change' calculated for a particular model is derived from the bias-corrected future and historical data from that model (not with the ERA-5 reanalysis). A sign test is used in order to identify statistically significant changes; the projected change in a given grid cell is statistically significant if the sign of the projected changes is the same across at least 8 of the 10 models. This corresponds to a $p$ value of 0.055. The sign test is selected since it is a robust statistical test even for small samples sizes, such as in this case where $N=10$.

\section{Results \& discussion}
\label{sec:results}

\subsection{Mapping wake effects}
\label{sec:results_mapping_wake_effects}

\begin{figure}
    \centering
    \includegraphics[width=\textwidth]{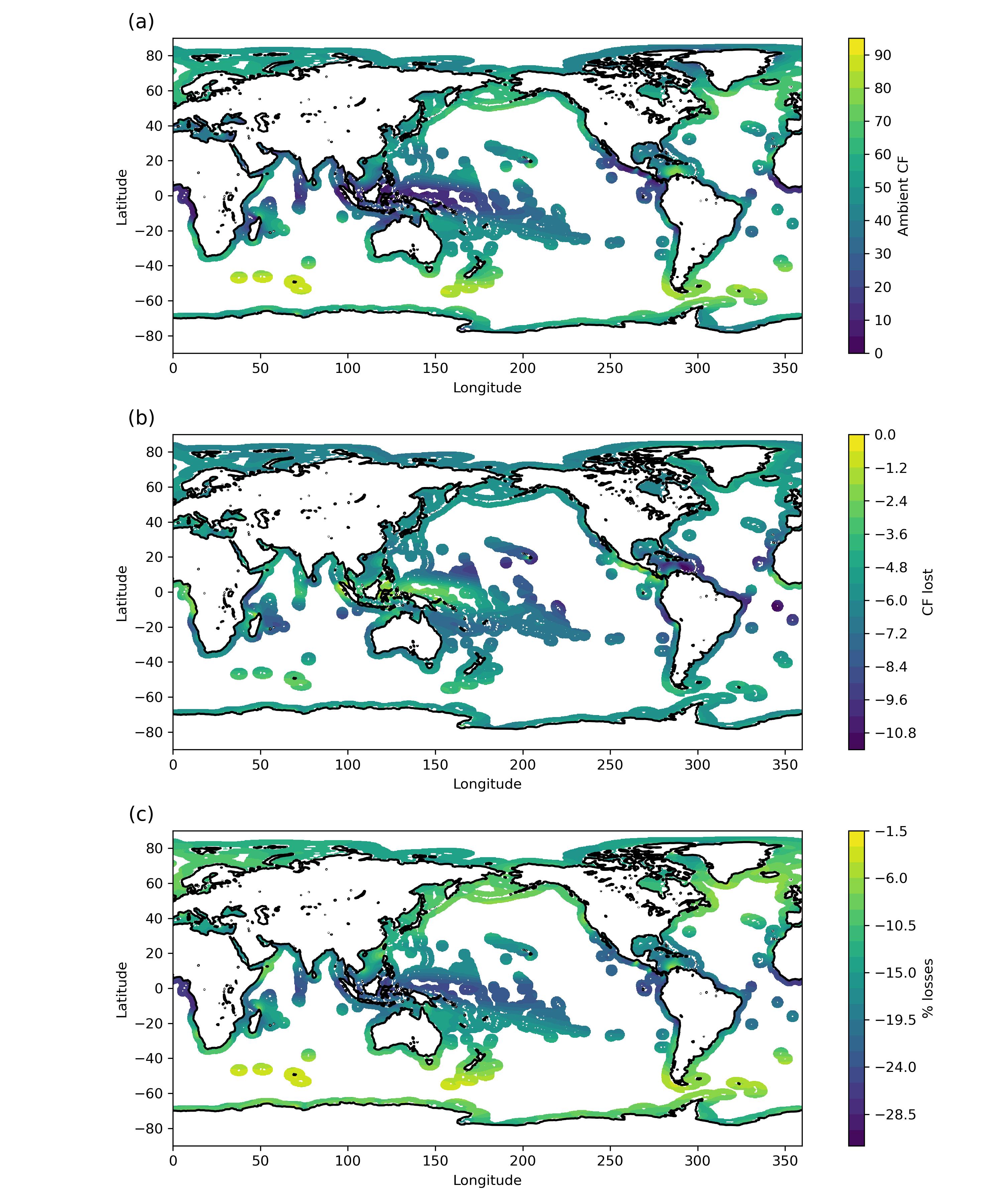}
    \caption{(a): Ambient capacity factor within all EEZs globally. (b): Capacity factor lost to intra-farm wakes, given by equation \eqref{eq:cf_lost}. (c): \% losses, defined by equation \eqref{eq:pc_power_lost}.}
    \label{fig:global_era5}
\end{figure}

Figure \ref{fig:global_era5}(a) shows the global distribution of ambient capacity factors, for the 10 \si{\mega\watt} reference turbine, based on ERA-5 data from our `historical' period of 1995--2014. Consistent with well-established global wind climatology \citep{zheng2018rezoning}, winds are strongest in the mid-latitudes, and in particular the results show very high ambient capacity factors for latitudes between -40 and -60. Wind potentials are smallest in the tropics. This result constitutes the most basic form of wind potential mapping, where focused only in the mean ambient wind power.

Figure \ref{fig:global_era5}(b) shows the global distribution of the capacity factor lost to intra-farm wakes for the 10 by 10 reference array averaged over 36 orientations, given by equation \eqref{eq:cf_lost}. Figure \ref{fig:global_era5}(c) shows the \% losses given by equation \eqref{eq:pc_power_lost}. Comparing with figure \ref{fig:global_era5}(a), in general, regions of high (low) ambient potential coincide with \% losses of low (high) magnitude. Since wake losses are negative, this amounts to a positive correlation between ambient CF and \% losses.

\begin{figure}
    \centering
    \includegraphics[width=0.45\textwidth]{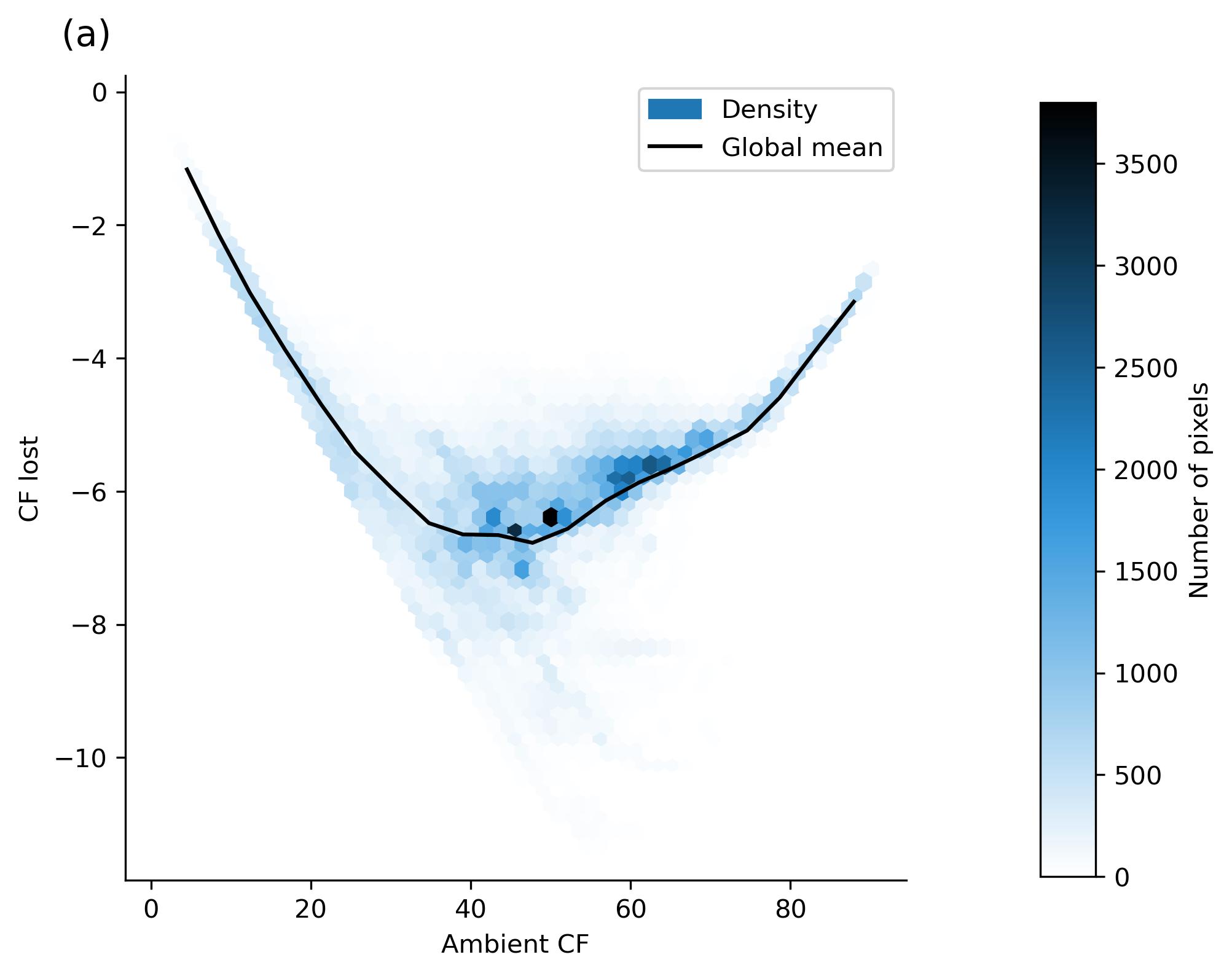}%
    \includegraphics[width=0.45\textwidth]{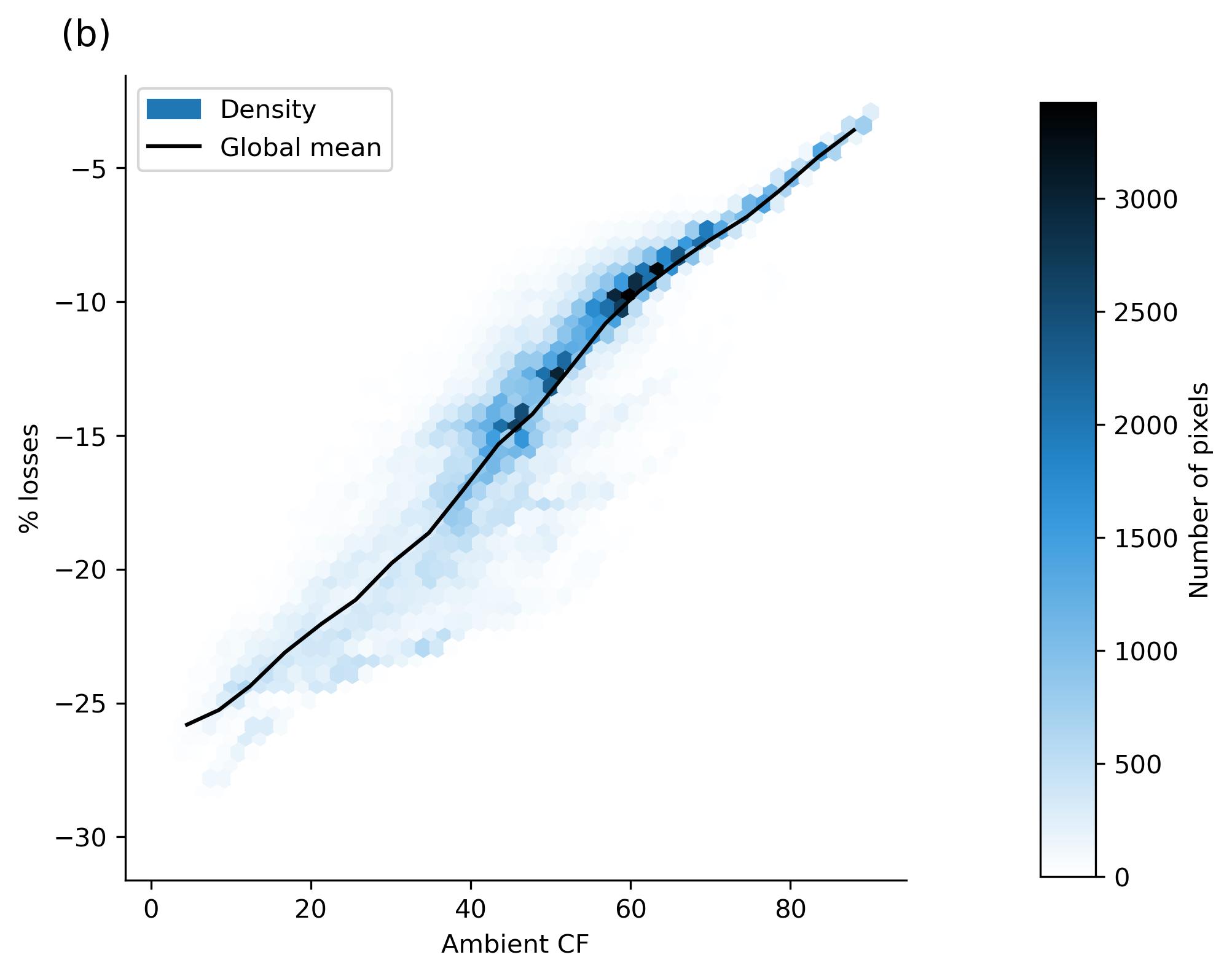}%
    \caption{(a) CF lost, and (b) \% power losses, due to intra-farm wakes, as a function of ambient CF. Solid lines show the global mean losses as a function of ambient CF. Figure (a) shows that the greatest absolute losses in power are found in regions where the ambient CF is around 40\%. Figure (b) indicates a broadly positive correlation between \% losses and ambient CF (although note that losses are defined as negative). There is also a significant spread in losses for a given ambient CF, particularly for ambient CFs in the range 40--60\%.}
    \label{fig:global_cf_lost_vs_bg_cf}
\end{figure}

Investigating this relationship further, figure \ref{fig:global_cf_lost_vs_bg_cf} shows the relationship between the ambient CF and (a) CF lost to wakes, and (b) \% wake losses. The solid lines show the global mean losses, as a function of ambient CF. Focusing on figure \ref{fig:global_cf_lost_vs_bg_cf}(a), for low ambient CFs, there is a near-linear relationship. However, for moderate ambient CFs of around 30\% or higher, there is a bifurcation. For most locations, the CF lost begins to reduce in magnitude (closer to zero) with increasing ambient CF. This can be explained by the decreasing fraction of wind speeds which fall between the cut-in and rated speeds of our reference wind turbine (3 and 11 \si{\metre\per\second} respectively), where wake effects are most significant. Most regions with higher average wind speeds produce greater ambient power, and also spend less time in this wake-sensitive interval, resulting in a reduction in absolute CF lost. However, there are some regions for which the near-linear relationship between ambient CF and CF lost continues, resulting in much greater wake losses. Looking at regions with ambient CF of, for example, 55--60\%, there is a very broad spread of CF lost, and hence net CF which is extracted by our `standard' farm.

\begin{figure}
    \centering
    \includegraphics[width=0.9\textwidth]{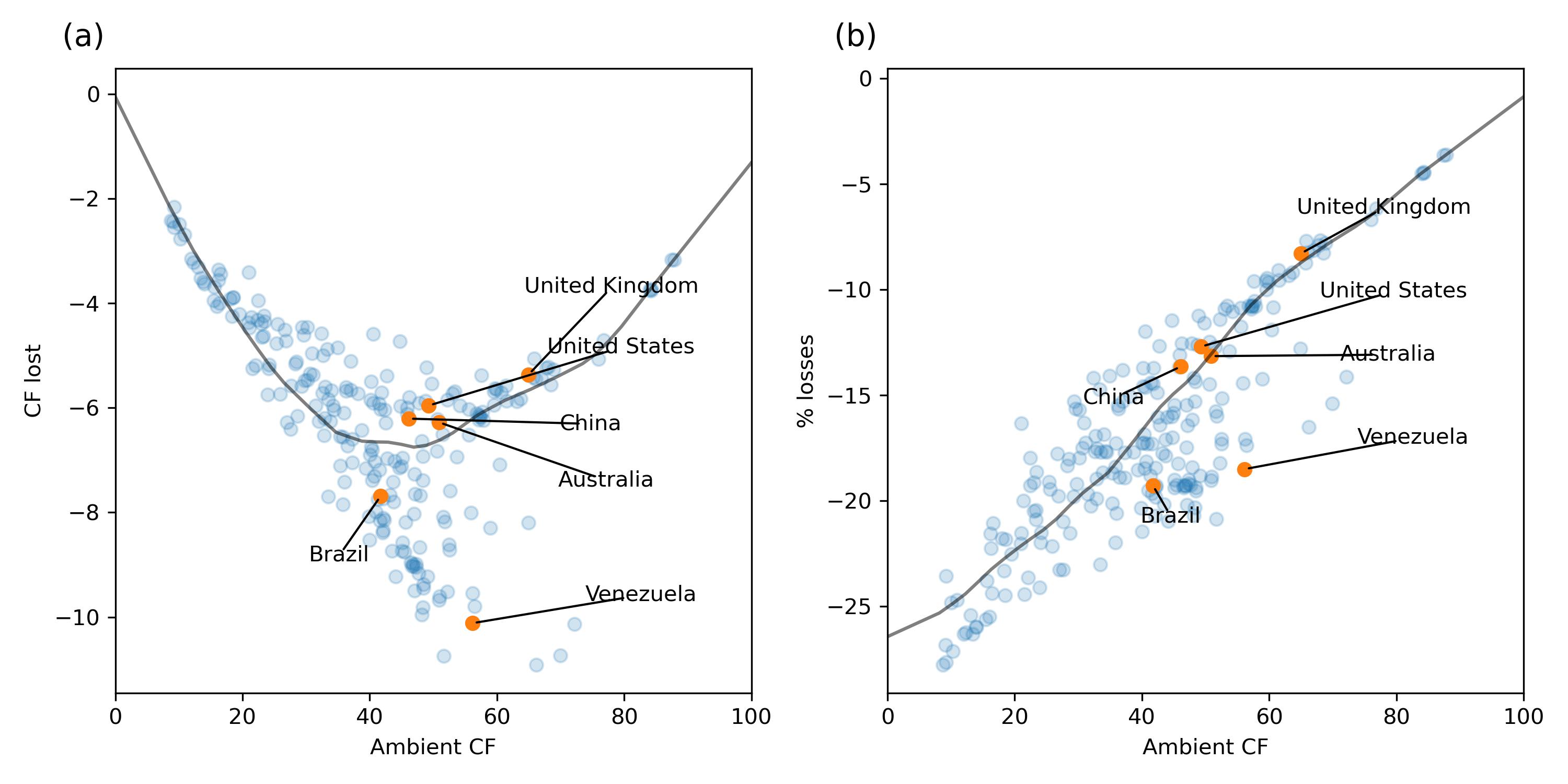}
    \caption{Average (a) CF lost, and (b) \% power losses, versus ambient CF, aggregated per country. Lines show global mean.}
    \label{fig:avg_per_country}
\end{figure}

Figure \ref{fig:avg_per_country} reproduces the results of figure \ref{fig:global_cf_lost_vs_bg_cf}, but presents results aggregated per country, with selected countries highlighted. The United Kingdom, with access to high-resource offshore regions including in the North Sea and Irish Sea, has a high mean ambient CF of 65\%, and relatively modest \% wake-induced losses of 8.3\%. The United States, Australia and China have moderate ambient CF, and also moderate \% losses of around 13\%. Venezuela is included as a fairly extreme example of high wake losses which is fairly typical of Atlantic South America; it has a strong ambient potential of 56\%, but suffers more than almost any other country in terms of absolute wake-induced losses, with 10.1\% CF lost, equivalent to 18.5\% losses. At 41.7\%, Brazil has the lowest ambient CF of the highlighted countries, and also suffers from greater than average \% losses of 19.3\%.

\begin{figure}
    \centering
    \includegraphics[width=\textwidth]{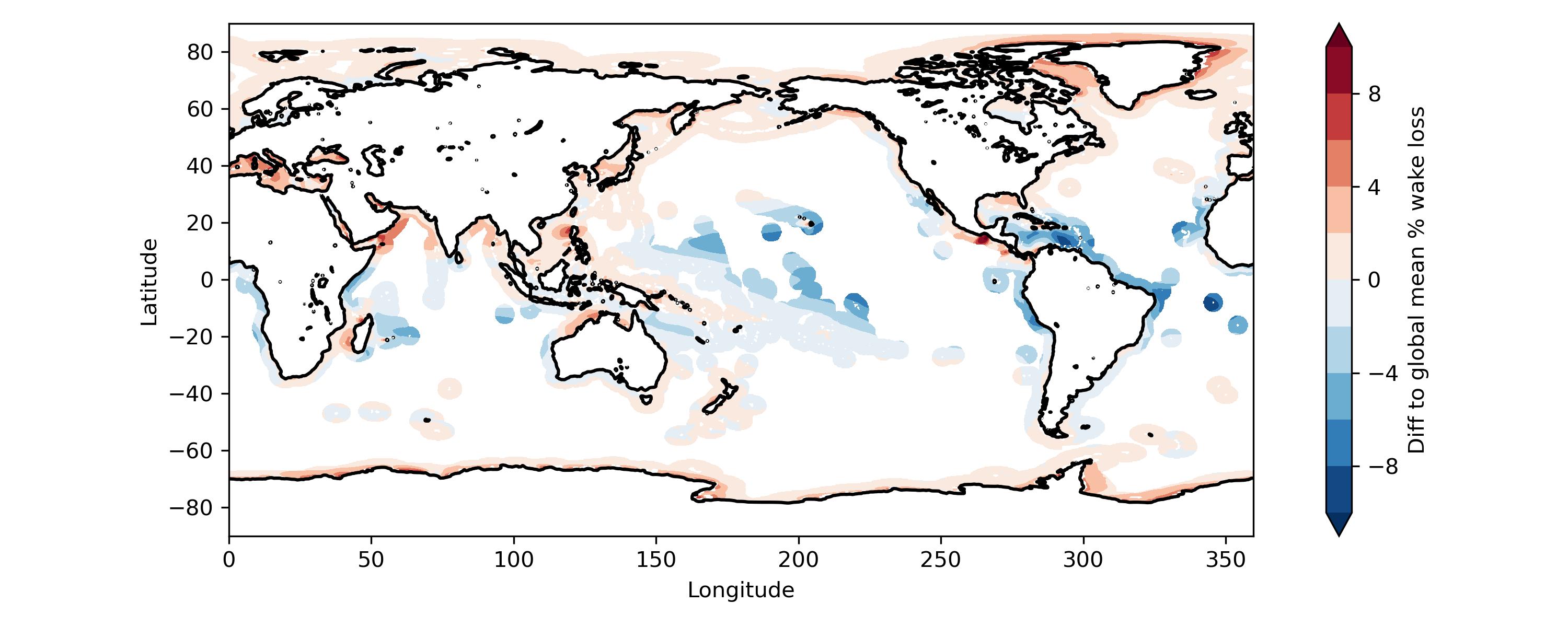}
    \caption{Difference between actual \% loss and global mean \% losses based on the fitted function in figure \ref{fig:global_cf_lost_vs_bg_cf}(b). Further detail for specific regions is shown in figure \ref{fig:diff_to_mean_by_country}. Broadly, larger wake losses (negative values) are found nearer the equator.}
    \label{fig:diff_to_mean_era5}
\end{figure}

\begin{figure}
    \centering
    \includegraphics[width=0.95\textwidth]{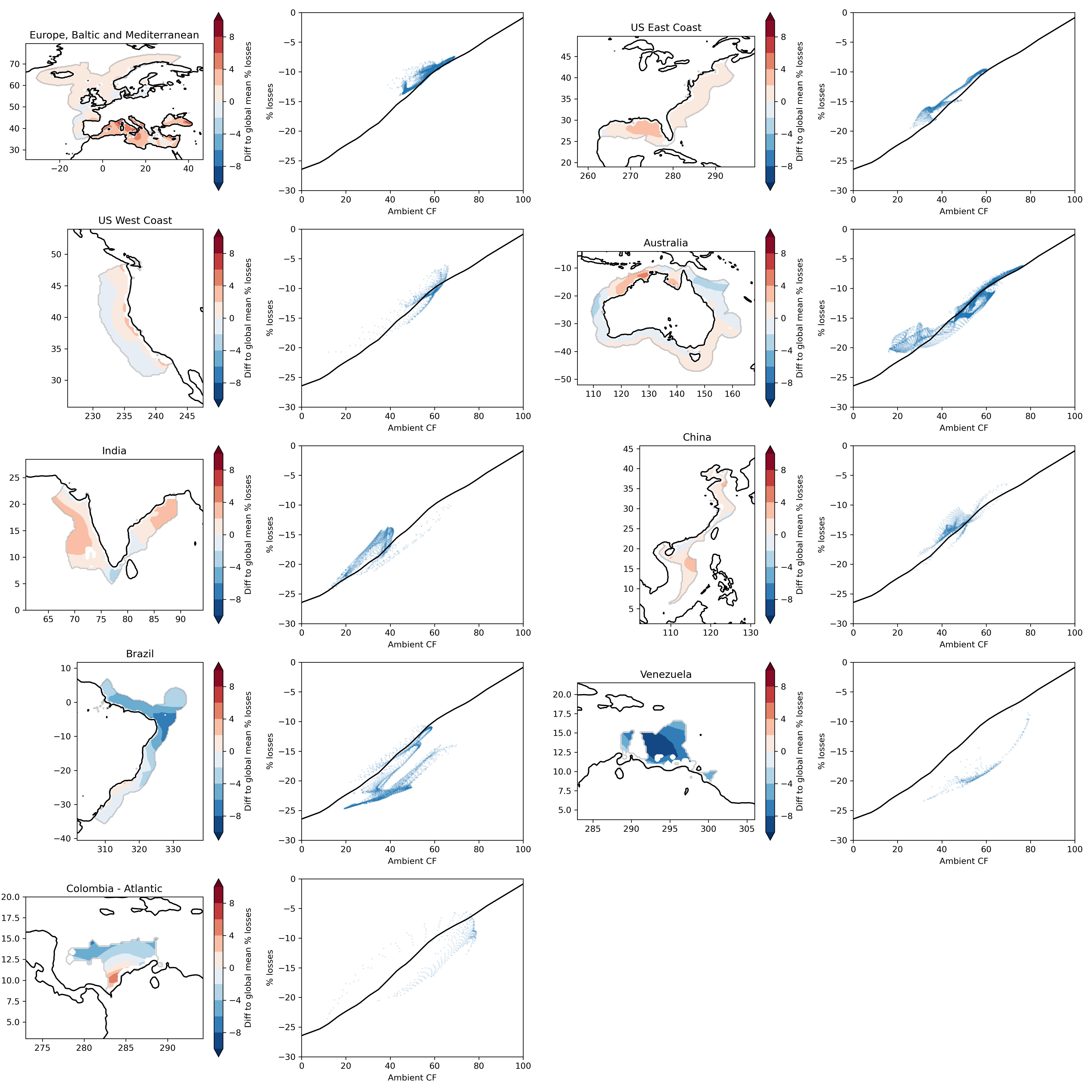}
    \caption{Ambient CF, \% losses and differences to global mean \% losses, for selected regions and countries.}
    \label{fig:diff_to_mean_by_country}
\end{figure}

Figure \ref{fig:diff_to_mean_era5} presents the difference between the actual \% losses, and the `global mean' \% losses as a function of ambient CF, shown by the solid line in figures \ref{fig:global_cf_lost_vs_bg_cf}(b) and \ref{fig:avg_per_country}(b). In general, the mid-latitudes suffer from smaller than average wake-induced losses (hence differences to the mean are positive), while the tropics suffer from larger losses (negative differences to the mean). It should be emphasised that these losses are relative to the global mean for the local ambient CF. Selected regions of interest are highlighted in figure \ref{fig:diff_to_mean_by_country}. Europe, including the Baltic and Mediterranean seas, possesses strong ambient potential, and favourable wake-induced losses. The US East Coast has moderate ambient potential and also has favourable losses. The US West Coast has stronger ambient potential than the East Coast. Near-shore regions exhibit favourable losses, but for regions further offshore the losses are stronger. Australia, with such a large spatial footprint, has a very broad range of ambient CFs, although some of the higher-resource regions coincide with larger-than-average wake-induced losses. India has fairly low ambient potential but mostly favourable wake losses. China has moderate ambient CFs, with slightly smaller-than-average wake effects. Atlantic South America suffers from some of the worst wake-induced losses. As identified in figure \ref{fig:avg_per_country}, Venezuela has strong ambient potential but very large wake-induced losses. Brazil similarly suffers from significant losses, as does most of the Atlantic coast of Colombia. These two regions were considered `markets to watch' by \citet{gwec_2023}; these results suggest that a different approach to planning and design may be required, compared with Europe, due to the stronger wake effects experienced.

Overall, these results reveal the inadequacy of any (global) resource assessment which neglects wake effects, or which assumes spatially uniform losses. Neither absolute or percentage losses are spatially uniform. Both exhibit overall trends with respect to ambient CF, but there remains significant variation even among locations with the same ambient wind potential. In terms of absolute losses, there is a turning point where regions with greater ambient potential actually suffer from smaller absolute wake-induced losses; this pertains mostly to middle-latitude regions. In contrast, tropical regions with moderate to high ambient potential are likely to suffer from the largest wake-induced losses, with tropical South America particularly impacted.

\subsubsection{Explaining the spatial variation in wake-induced losses}
\label{sec:results_explaining_variation}

\begin{figure}
    \centering
    \includegraphics[width=0.5\textwidth]{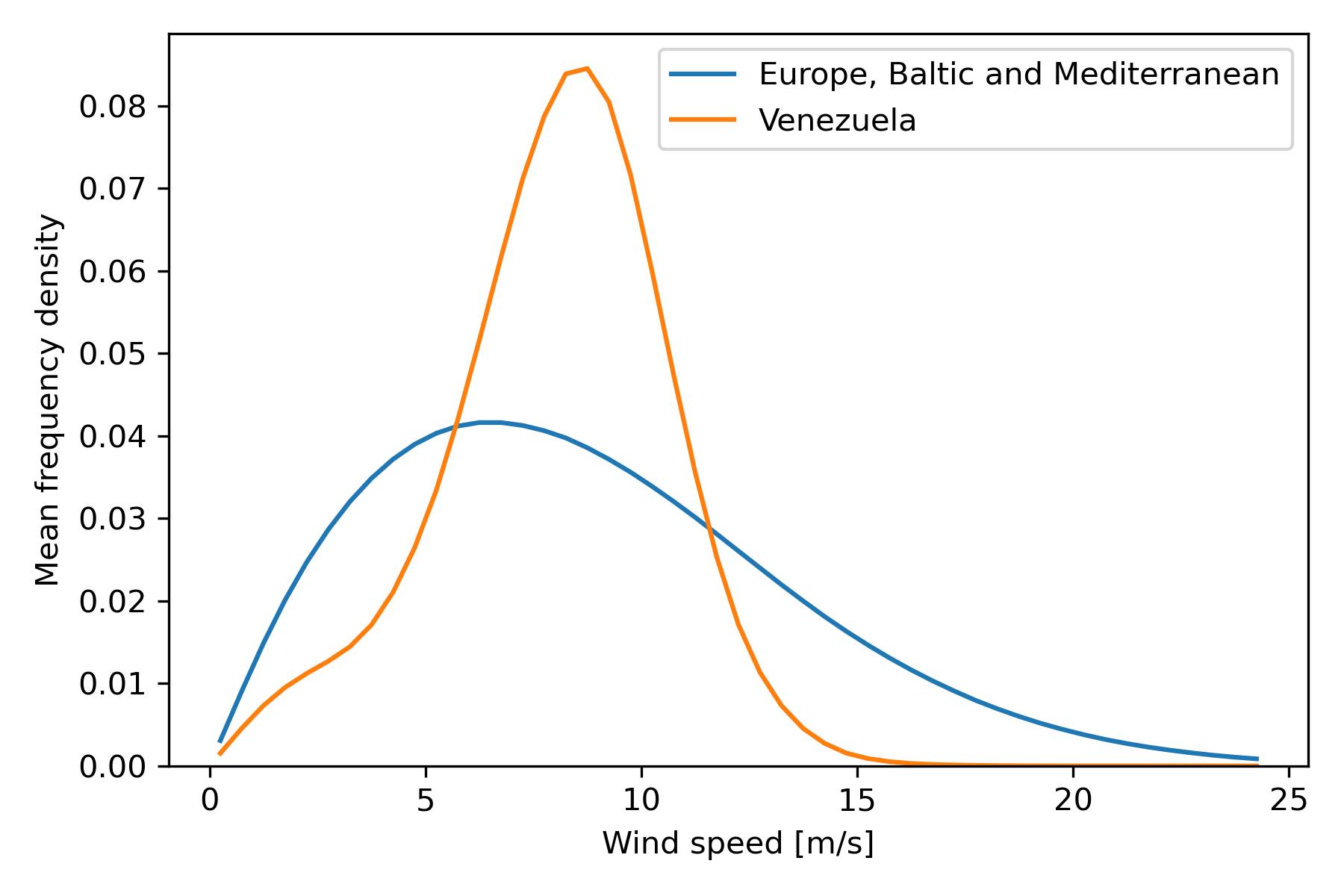}
    \caption{Mean wind speed distributions for Europe and Venezuela.}
    \label{fig:wind_speed_dists}
\end{figure}

\begin{figure}
    \centering
    \includegraphics[width=0.6\linewidth]{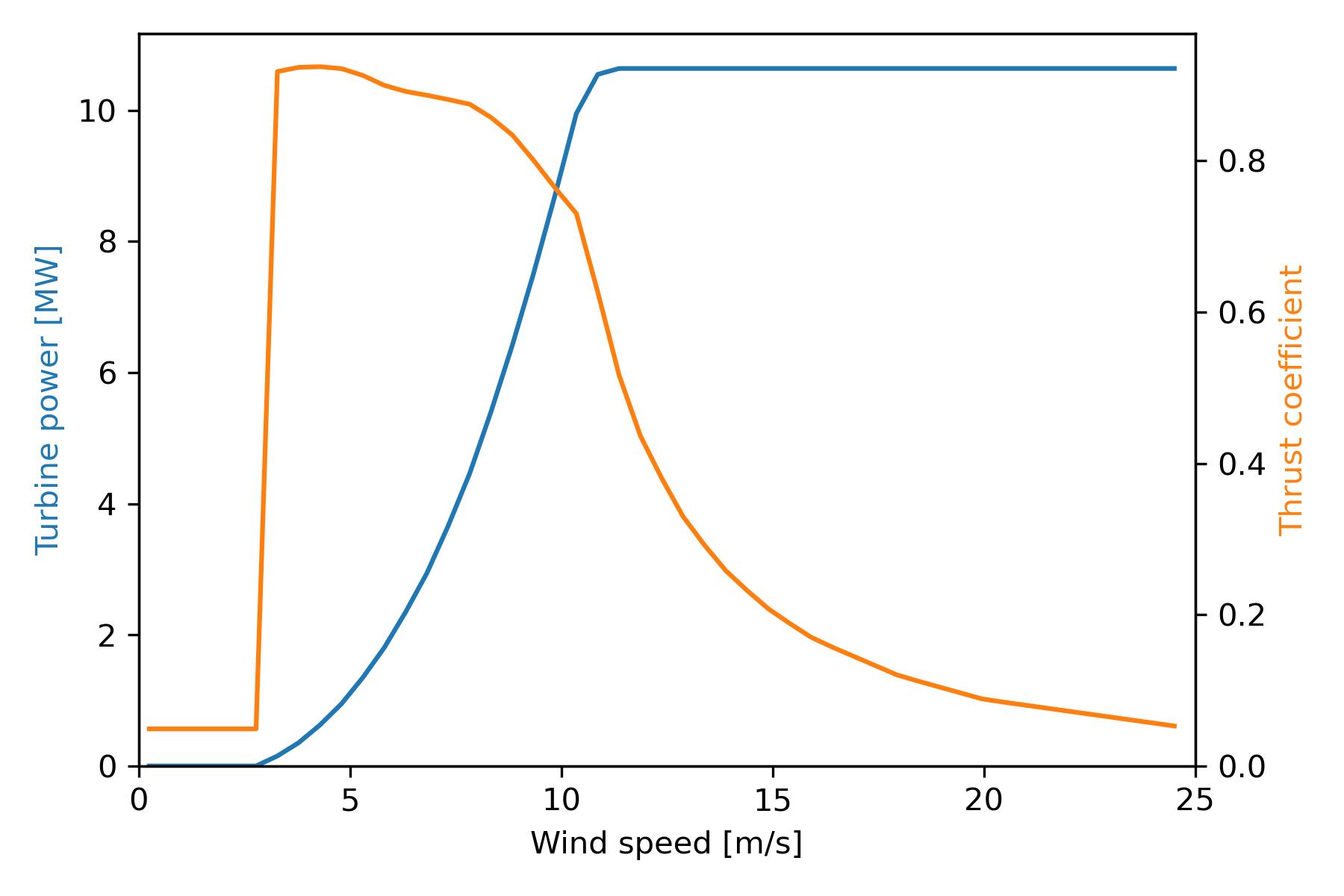}
    \caption{Power and thrust curves for the 10 \si{\mega\watt} reference turbine used within this work \citep{bortolotti2019iea}.}
    \label{fig:power_thrust}
\end{figure}

To explain the observed spatial variation in wake-induced losses, it is necessary to inspect specific wind speed distributions, in comparison to the turbine power and thrust curves. Figure \ref{fig:wind_speed_dists} shows the mean wind speed distributions for Europe (including the Baltic and Mediterranean seas), and Venezuela, which was identified above as a country that experiences very high losses. The ambient capacity factors for these regions are very similar (55.8\% for Europe, 56.1\% for Venezuela), but the CF lost to wakes differs significantly (-5.5\% for Europe, -10.1\% for Venezuela). This is due to their differing wind speed distributions, as shown in figure \ref{fig:wind_speed_dists}. Venezuela exhibits a far narrower wind speed distribution than Europe, coinciding with the most wake-sensitive part of the turbine power and thrust curves (figure \ref{fig:power_thrust}), which is between around 3 and 11 \si{\metre\per\second}. In Europe, a substantial part of the wind speed distribution sits above 11 \si{\metre\per\second}, where power output is at its maximum and is insensitive to wake effects.

\begin{figure}
    \centering
    \includegraphics[width=0.55\textwidth]{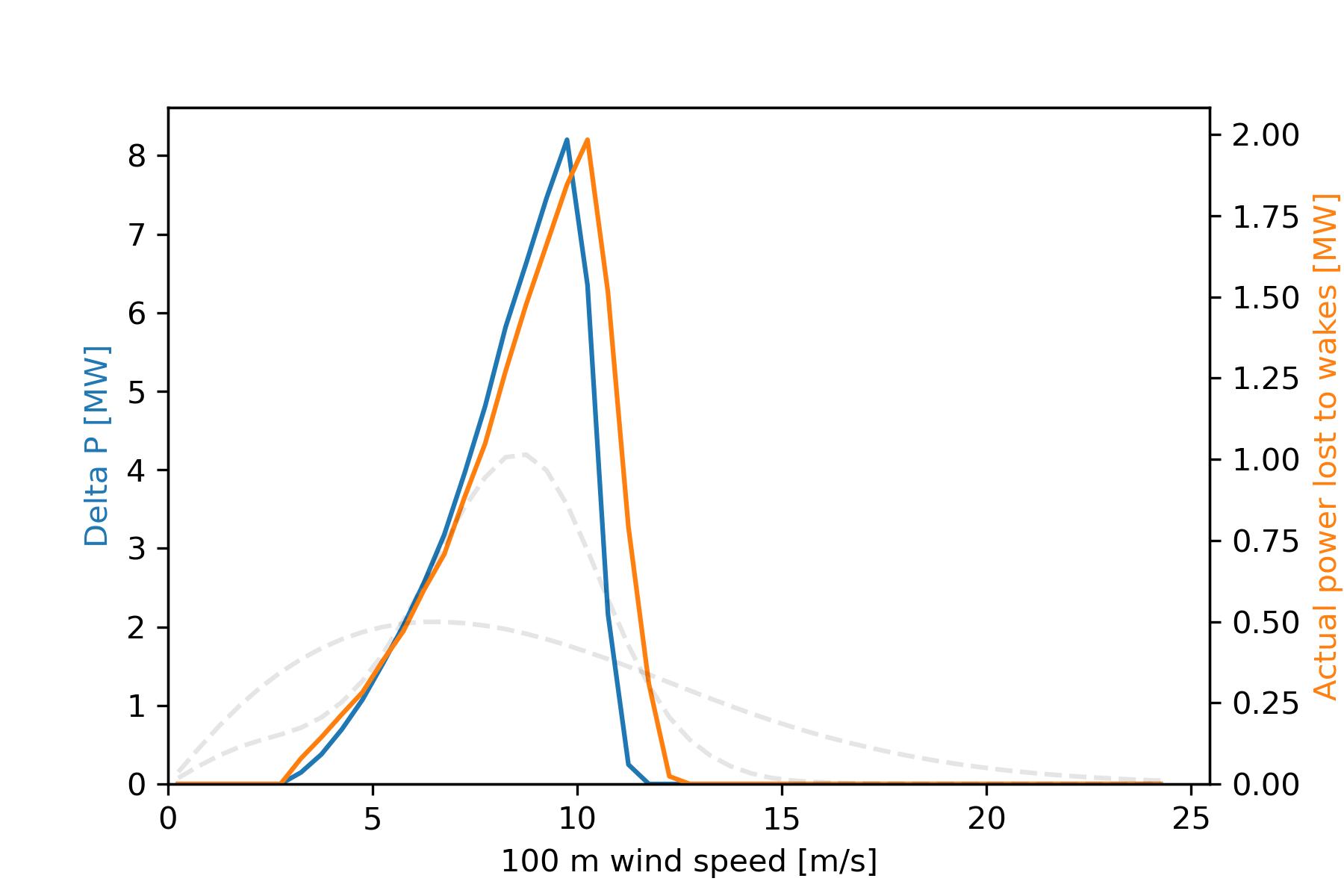}
    \caption{$\Delta P$ metric, given by equation \eqref{eq:delta_p}. This approximates the theoretical power lost by a downstream turbine, for a given upstream wind speed. The secondary $y$ axis shows the power loss by the `standard' farm, as a function of wind speed, as simulated using PyWake. The region of wind speeds with greatest wake effects is between the cut-in and rated speeds, or around 3 -- 11 \si{\metre\per\second}. The dotted lines indicate the wind speed distributions for Europe and Venezuela, from figure \ref{fig:wind_speed_dists}, highlighting the significant overlap between the wind speeds for Venezuela, and the $\Delta P$ metric.}
    \label{fig:delta_p}
\end{figure}

This influence of the wind speed distribution on wake effects can be more formally explained by considering one-dimensional momentum theory (see \eg \citet{kulunk2011aerodynamics}). The power coefficient of a turbine $C_p$ can be related to the axial induction factor $a$, by
\begin{equation}
    C_p = 4a(1-a)^2.
\end{equation}
Given the wind turbine power curve data for our selected turbine model (see figure \ref{fig:power_thrust}), $a$ can be derived as a function of upstream wind speed.  Neglecting wake expansion, the reduction in wind speed downstream of a wind turbine is then given by
\begin{equation}
    U_w = U_\infty (1-2a),
\end{equation}
where $U_w$ is the wind speed in the wake, and $U_\infty$ is the upstream wind speed. The reduction in wind speed is therefore given by
\begin{equation}
    \Delta U = -2 a U_\infty.
\end{equation}
Given the turbine power curve $P(U)$, the reduction in power at a downstream turbine can then be estimated as
\begin{equation}
    \label{eq:delta_p}
    \Delta P(U_\infty) = \left.\frac{\text{d} P}{\text{d} U}\right|_{U_\infty} \Delta U = -2a \left.\frac{\text{d} P}{\text{d} U}\right|_{U_\infty} U_\infty.
\end{equation}
Figure \ref{fig:delta_p} shows this $\Delta P$ estimate as a function of wind speed. This is consistent with our earlier statements regarding the most important wind speed interval for wake effects, between around 3 and 11 \si{\metre\per\second}. Figure \ref{fig:delta_p} also shows the actual average power lost per turbine by the `standard' farm studied within this work, as a function of wind speed, as simulated using PyWake. Although $\Delta P(s)$ corresponds to the theoretical power lost by a single turbine downstream of another, neglecting wake expansion, it is clear that it is a good indicator of wake-induced losses at a given wind speed. There appears to be a slight offset in the position of the peak losses, which we attribute to the simplicity of the estimates derived in this section, and in particular the first-order approximation of equation \eqref{eq:delta_p}, which breaks down around rated speeds where the power curve is highly nonlinear.

To derive a theoretical estimator for the overall mean wake-induced losses, this $\Delta P(s)$ is combined with the wind speed distribution at a given location $f_s$, according to
\begin{equation}
    \label{eq:loss_estimator}
    \sum_s \Delta P(s) f_s.
\end{equation}
This quantity is very strongly correlated with the simulated wake losses, with an $R^2$ coefficient of 0.96.

It is important to note that this formulation does not depend on the total capacity or layout of the wind farm, but only on the power and thrust curves of the selected turbine. It follows that turbine models with different thrust and power curves will produce a different spatial pattern of wake-induced losses. It may further be possible to mitigate losses by careful selection of turbine types appropriate to the wind speed distribution at the site where they will be deployed.
Further exploration of the wake response for different turbine types is beyond the scope of the present study but will be considered in future work.

\subsection{Mapping layout optimisation potential}
\label{sec:results_opt_potential}

\begin{figure}
    \centering
    \includegraphics[width=\textwidth]{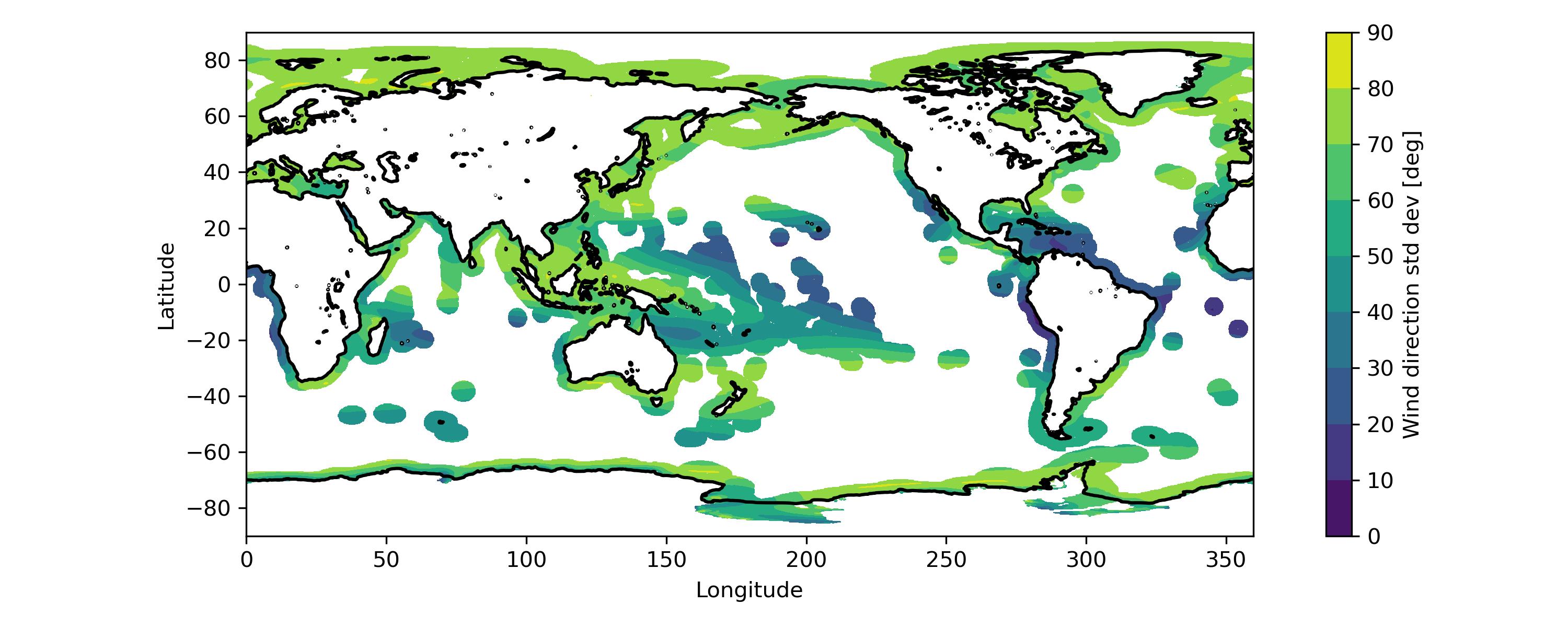}
    \caption{Circular standard deviation of wind direction. The wind direction distribution is narrowest nearer the equator.}
    \label{fig:wind_dir_std_dev}
\end{figure}

Figure \ref{fig:wind_dir_std_dev} shows the global distribution of the circular standard deviation of the wind direction. The narrowest wind direction distributions are found near the equator, where the wind climate is dictated by the trade winds. Mid-latitudes experience a broader spread of wind directions.

\begin{figure}
    \centering
    \includegraphics[width=\textwidth]{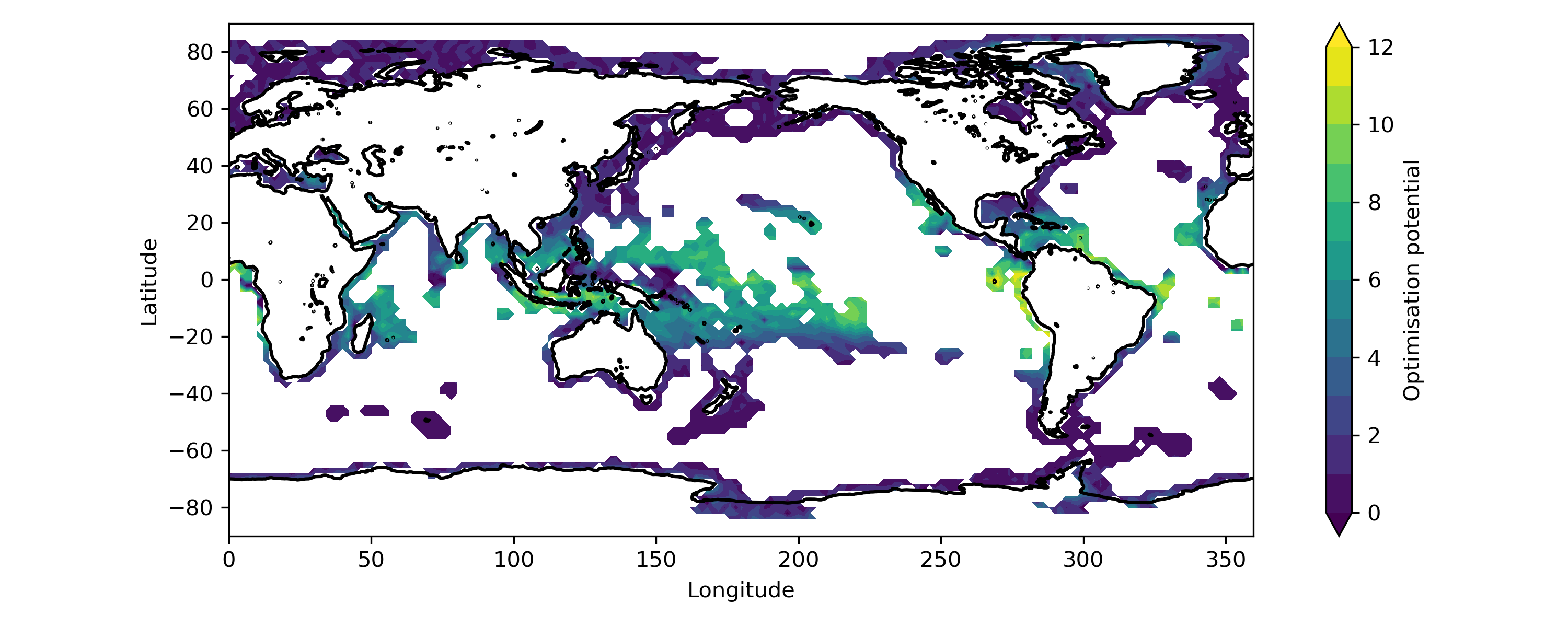}
    \caption{Optimisation potential. The greatest optimisation potentials are found in regions with the narrowest wind direction distributions, which tend to be found nearer the equator; see figure \ref{fig:wind_dir_std_dev}.}
    \label{fig:optimisation_potential}
\end{figure}

Figure \ref{fig:optimisation_potential} shows the global distribution of the array layout optimisation potential, which is defined as the percentage improvement in mean power which can be obtained by an optimised layout, compared with the initial square farm layout. There is a negative correlation between wind direction standard deviation and optimisation potential. That is, narrower wind speed distributions are more conducive to array layout optimisation. The greatest optimisation potentials are found in northern South America, particularly on the western (Pacific) coast.

\begin{figure}
    \centering
    \includegraphics[width=0.45\textwidth]{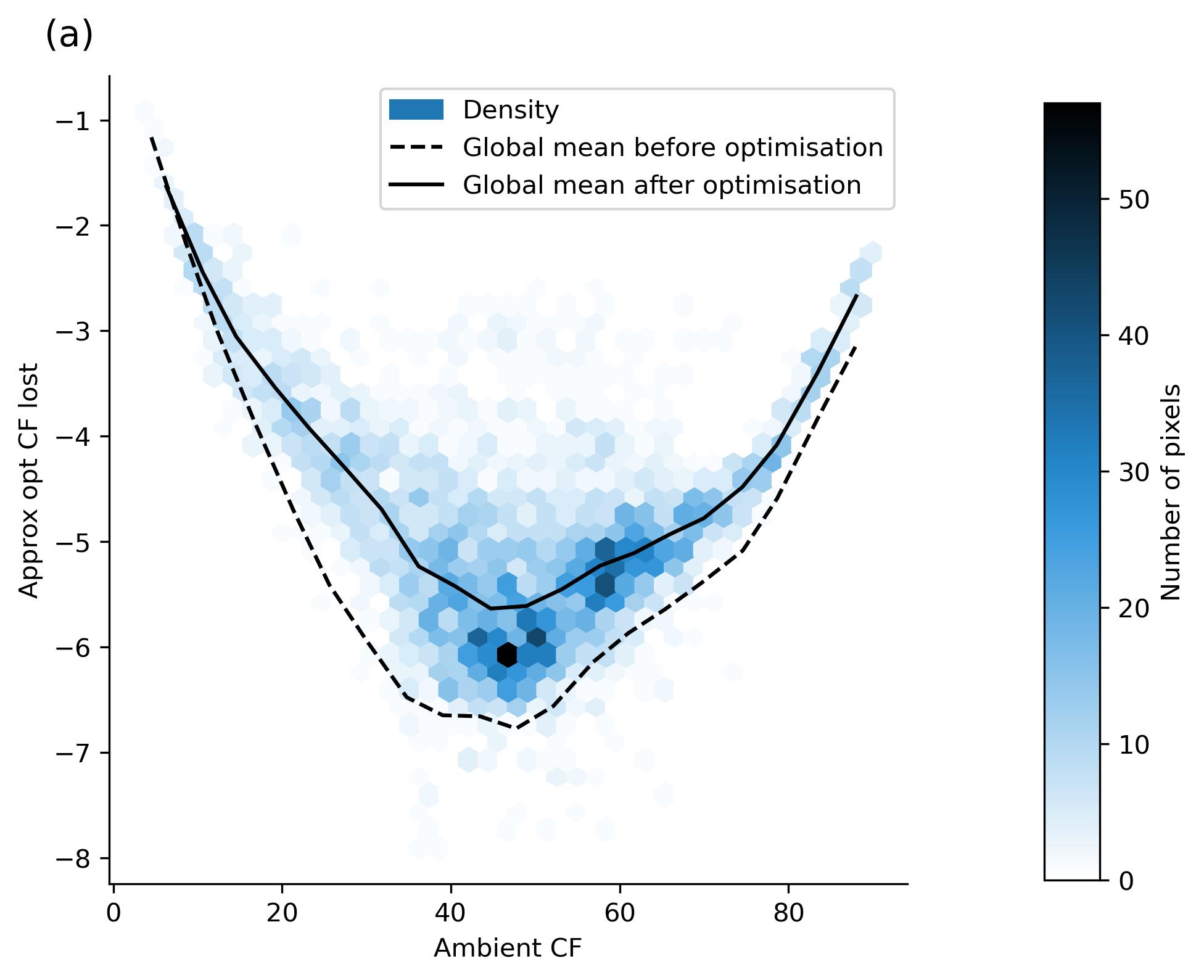}%
    \includegraphics[width=0.45\textwidth]{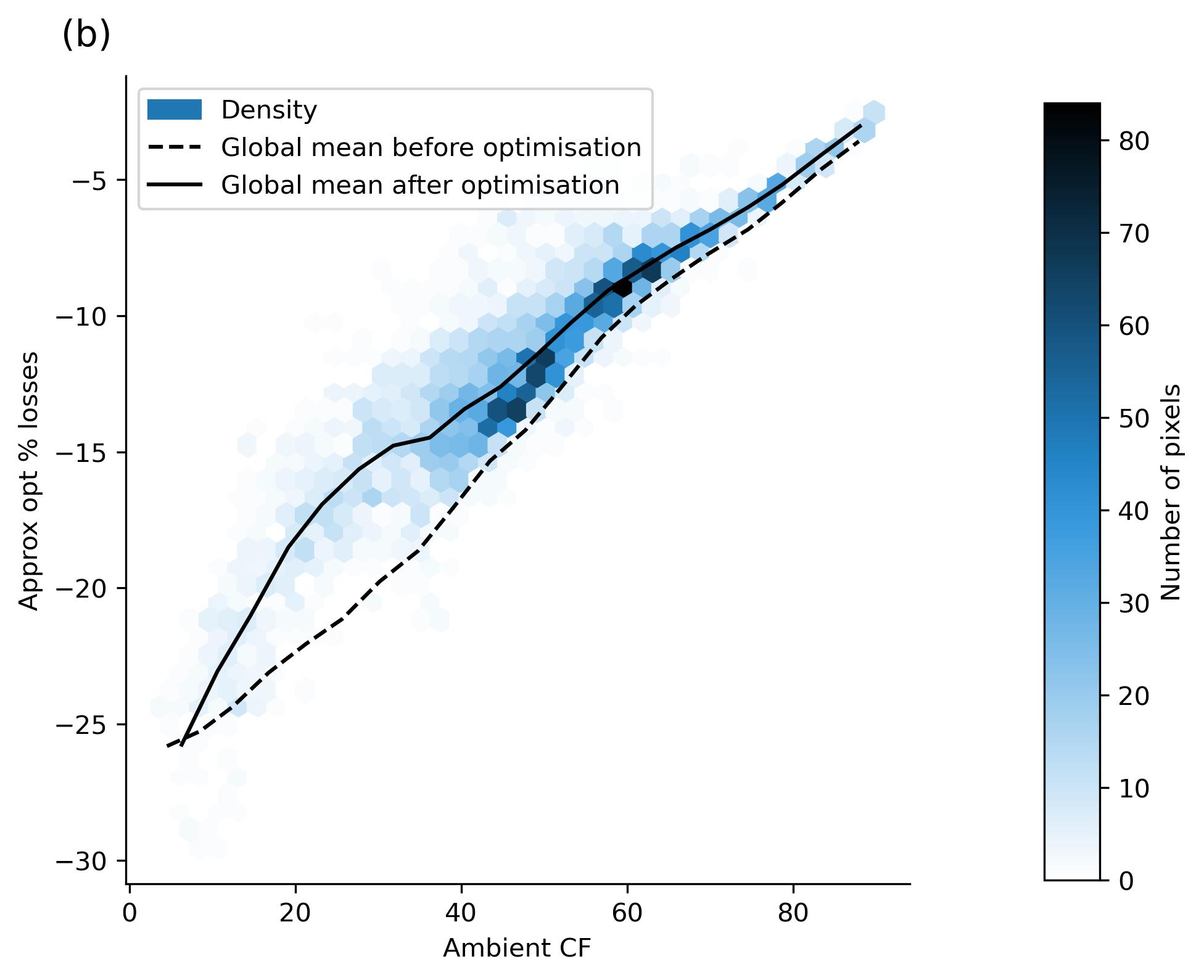}%
    \caption{(a) CF lost, and (b) \% power losses, due to intra-farm wakes, as a function of ambient capacity factor, after optimising wind farm layout. The dashed lines indicate the global means prior to optimisation, from figure \ref{fig:global_cf_lost_vs_bg_cf}.}
    \label{fig:global_cf_lost_vs_bg_cf_opt}
\end{figure}

Comparing the optimisation potential of figure \ref{fig:optimisation_potential} with the global distribution of excess wake losses of figure \ref{fig:diff_to_mean_era5}, the regions (mostly equatorial) where the wake losses are particularly high also have the greatest optimisation potential. This is also shown in figure \ref{fig:global_cf_lost_vs_bg_cf_opt}, which should be compared with figure \ref{fig:global_cf_lost_vs_bg_cf}. Mean losses at all ambient CFs are improved by optimisation. Looking at the absolute losses (figs \ref{fig:global_cf_lost_vs_bg_cf}(a) and \ref{fig:global_cf_lost_vs_bg_cf_opt}(a)), the optimisation has the greatest benefit for ambient CFs around 40--60\%. In terms of percentage losses (figs \ref{fig:global_cf_lost_vs_bg_cf}(b) and \ref{fig:global_cf_lost_vs_bg_cf_opt}(b)), the greatest improvements are found for ambient CFs of around 10--40\%.

Overall, these results reveal that array layout optimisation has the potential to mitigate wake losses, and that the mitigation potential is often greatest in regions where wake losses are most severe. As explained in section \ref{sec:results_explaining_variation}, the higher wake losses in regions such as Venezuela are due to the narrower wind speed distribution. The results of this section show that these regions are also associated with narrower wind direction distributions, leading to greater mitigation opportunities via array layout optimisation.

These results implicitly assume that the wind distribution which is assumed during array design is applicable across the lifetime of the farm. Alternatively, climate projections could be used as the basis for robust array design with respect to future changes in wind distributions. However, this is beyond the scope of this study and is left for future work.

\subsection{Mapping future changes due to climate}
\label{sec:results_climate}

\begin{figure}
    \centering
    \includegraphics[width=\textwidth]{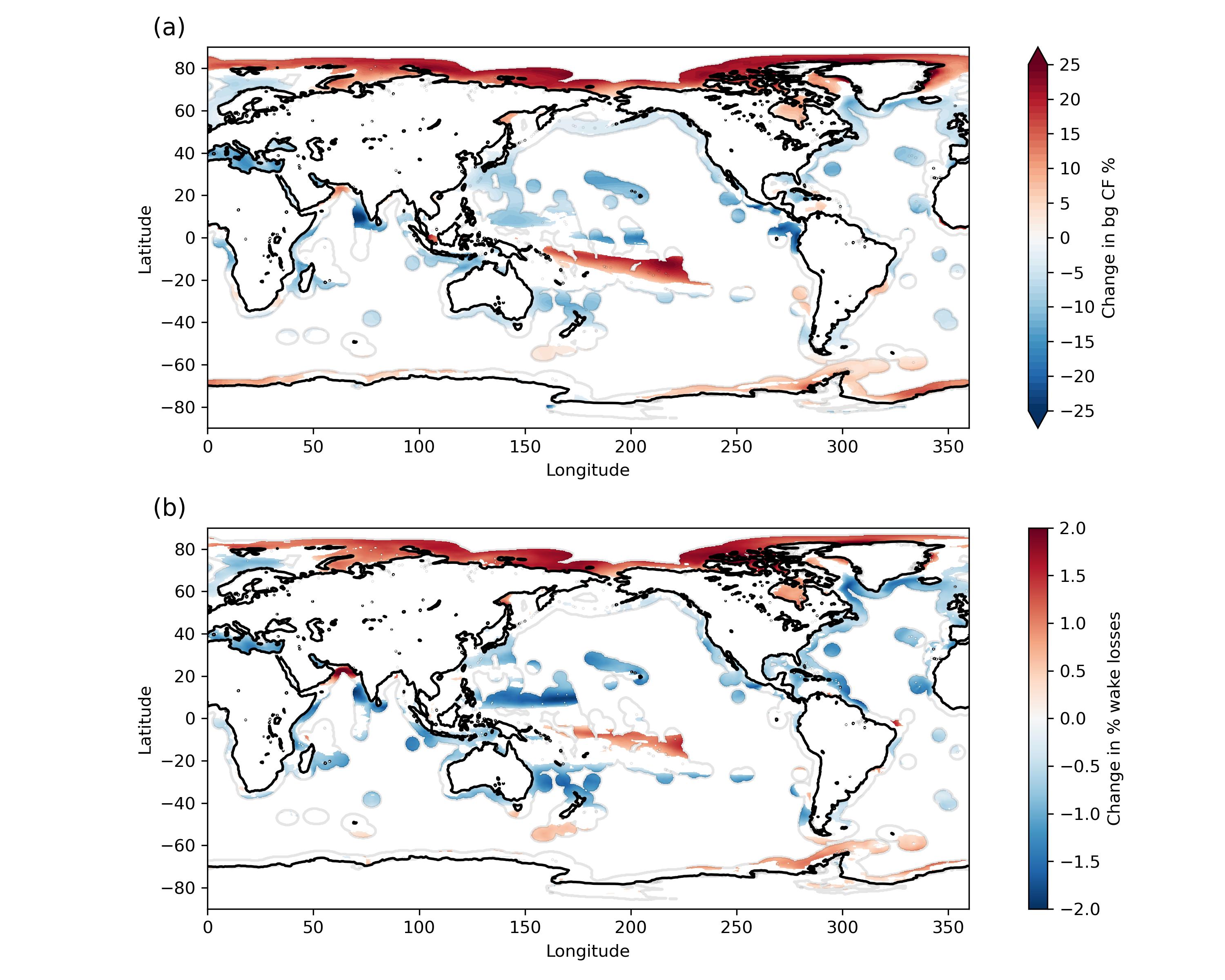}
    \caption{Mean projected changes over the 10 climate models in (a) ambient CF, (b) \% wake-induced losses, from the historical to the future scenarios. Only grid cells with statistically significant changes are shown. The \% change in power lost to wakes typically has the same sign as the \% change in ambient CF.}
    \label{fig:change_maps}
\end{figure}

Figure \ref{fig:change_maps}(a) shows the mean projected percentage change in ambient capacity factor across the 10 climate models, between the `historical' climate, and 2081--2100 under the SSP585 scenario. Only grid cells with a statistically significant change are shown, where our significance criterion is that at least 8 out of the 10 climate models show the same sign of change. The result is broadly consistent with the global analysis of \citet{ibarra2023cmip6}, as well as with other regional studies. Statistically significant decreases in offshore wind potential are observed across both Europe (consistent with \citep{carvalho2021wind}) and the US (consistent with \citet{martinez2022climate}). A decrease in potential is observed offshore of Northern China, which is consistent with \citet{zhang2021future,deng2024offshore}, although we do not find statistically significant increases in south China. Decreasing potentials are observed on the west coast of India \citep{basak2023foreseeing}, as well as most of Australia \citep{fournier2023impacts}. Our projections around South America are also consistent with the study of \citet{de2024assessment}.

\begin{figure}
    \centering
    \includegraphics[width=0.5\textwidth]{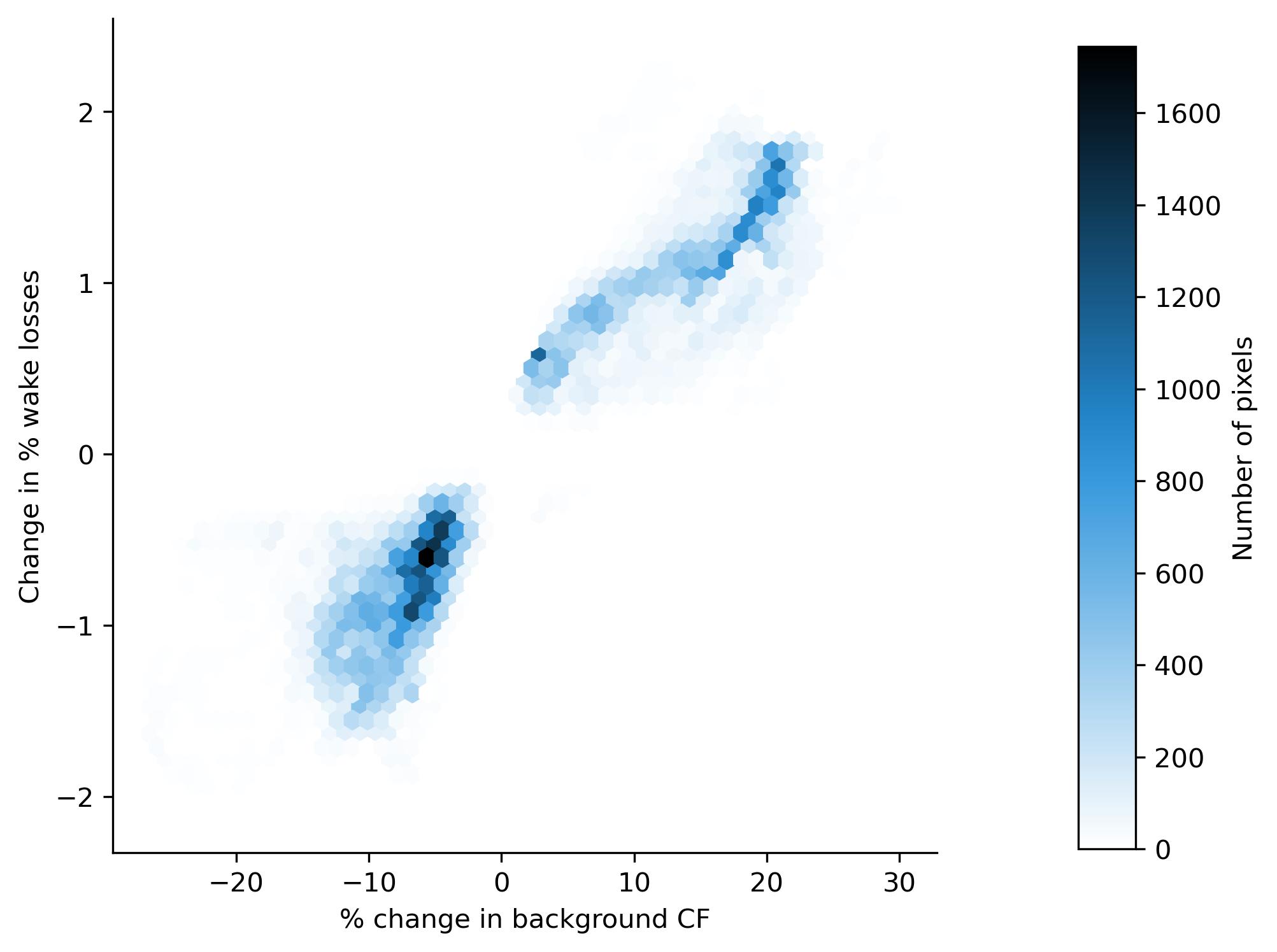}
    \caption{Change in \% wake losses versus \% change in ambient capacity factor, due to climate change. Only grid cells with significant changes are included.}
    \label{fig:change_scatter}
\end{figure}

\begin{figure}
    \centering
    \includegraphics[width=0.5\textwidth]{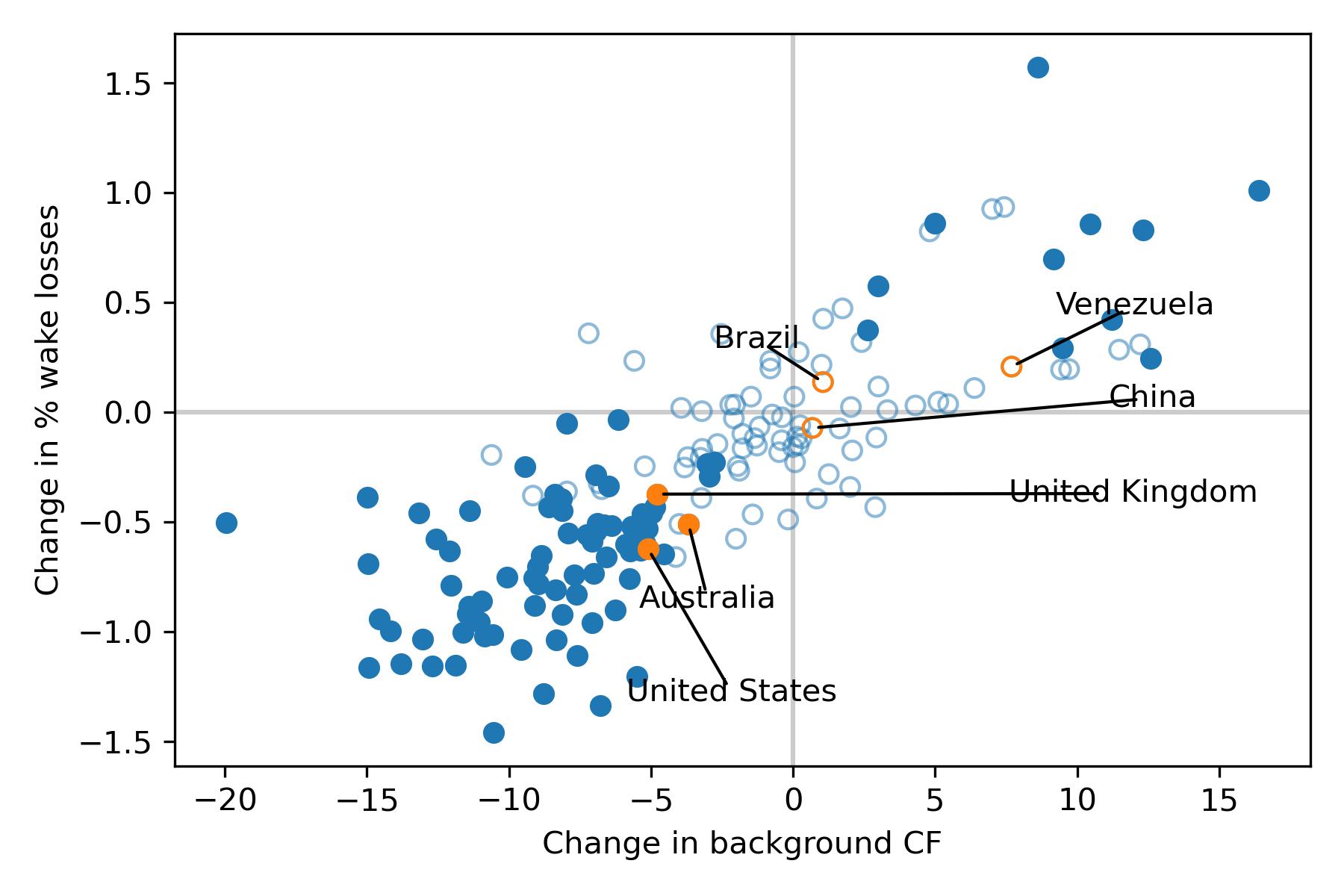}
    \caption{Change in \% wake losses versus \% change in ambient capacity factor, due to climate change, averaged by country. The mean is shown for all countries. Those with a statistically significant change are shown as solid points. Countries with insignificant changes are shown as hollow circles.}
    \label{fig:change_scatter_by_country}
\end{figure}

Figures \ref{fig:change_scatter} and \ref{fig:change_scatter_by_country} show the relationship between the \% projected change in ambient CF, and \% losses, per grid cell (figure \ref{fig:change_scatter}), and aggregated per country (figure \ref{fig:change_scatter_by_country}). For the aggregation by country, the significance is determined on a per-country basis, rather than by aggregation over only the significant grid cells belonging to each country. Both indicate a positive correlation between change in ambient CF, and change in \% wake losses. This is consistent with the trend found in section \ref{sec:results_mapping_wake_effects}, and in particular figure \ref{fig:global_cf_lost_vs_bg_cf}(b).

\begin{figure}
    \centering
    \includegraphics[width=\textwidth]{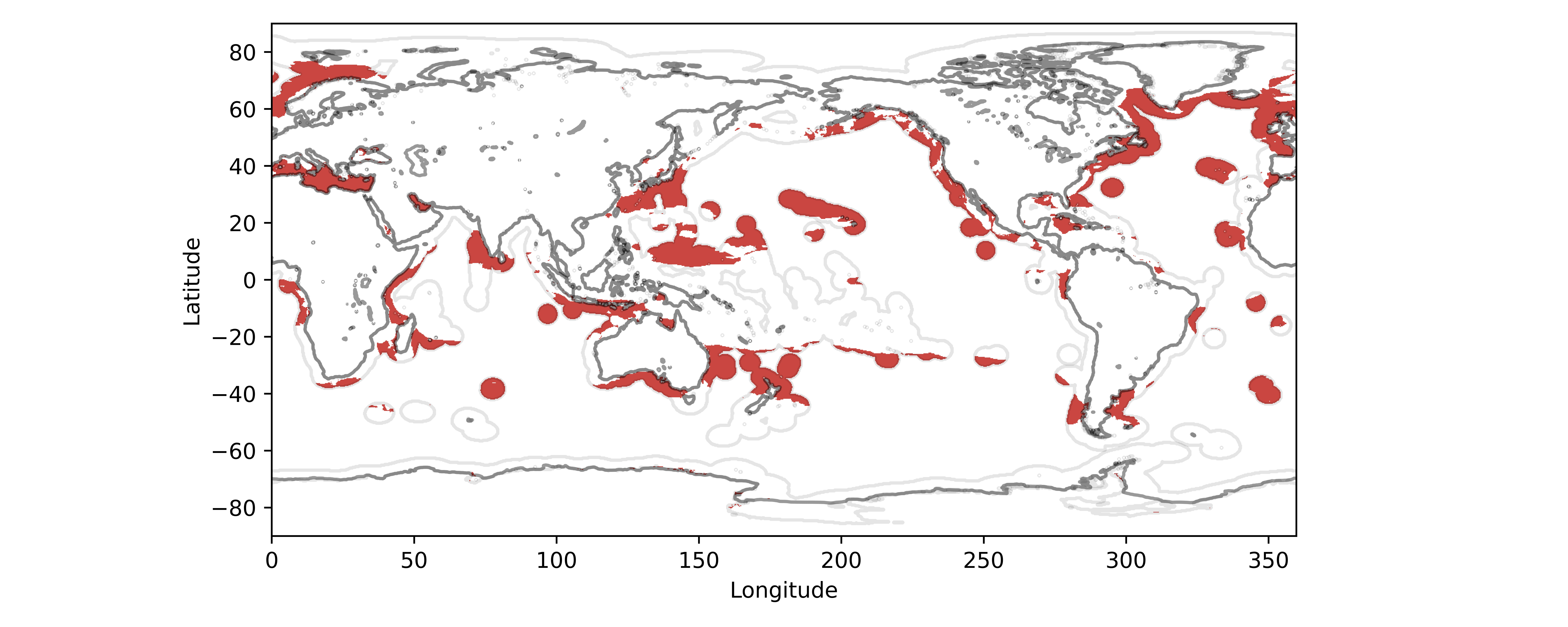}
    \caption{Highlighted regions indicate where ambient wind potential is projected to decrease, and percentage wake-induced losses are projected to increase in magnitude.}
    \label{fig:bg_down_wakes_down}
\end{figure}

The majority of grid cells/countries fall into the lower left quadrant of figures \ref{fig:change_scatter} and \ref{fig:change_scatter_by_country}, indicating that ambient potential decreases, but that percentage wake-induced losses increase in magnitude, exacerbating the climate change effect. Figure \ref{fig:bg_down_wakes_down} indicates the spatial distribution of these grid cells. These regions include the northern Atlantic, impacting the US, Atlantic Canada, Greenland, Iceland, Ireland and the UK, as well as the US west coast, parts of Australia, India and China, southern Chile, and a number of island states in the southern Pacific. In these regions, decreases in ambient wind potential due to climate change are exacerbated by an increase in the magnitude of the percentage wake-induced losses.

In contrast, countries falling into the upper right quadrant of figures \ref{fig:change_scatter} and \ref{fig:change_scatter_by_country} are projected to both increase in ambient potential and improve in wake losses; these countries include Russia and Canada (driven mostly by strong increases in resource in the Arctic), as well as other isolated regions such as Pakistan and Oman. There are no countries which fall within either the upper left or lower right quadrants for which the statistical significance threshold is met, although there are a small number of individual grid cells meeting these criteria, as shown in figure \ref{fig:change_scatter}.

Comparing the magnitudes of the various effects studied within this work, the direct impact of climate change on ambient power exceeds magnitudes of 25\% for some regions, such as India. The magnitudes of the baseline wake-induced losses (for the current climate) were similar, occasionally exceeding 25\% in regions of low ambient potential. Optimisation potentials reached up to 12\%, while the indirect climate change impact via wake effects has magnitudes of around 1-2\%. However, the relative size and significance of these effects varies by region. For example, there are some regions where optimisation potential exceeds the projected impact of climate change. This serves to further highlight that the challenges facing the offshore wind industry are highly region-dependent.

\section{Conclusions}
\label{sec:conclusions}

This study has investigated global spatial variation in (i) `ambient' offshore wind power potential, and (ii) intra-farm wake losses for a `standard' offshore wind farm with a design representative of near-future installations. The spatial distribution of ambient wind power potential follows well-established global wind climatology, with resources greatest in the mid-latitudes, and lowest in the tropics. There is a strong correlation between ambient resources and percentage wake-induced losses; regions with greater ambient resource tend to lose a smaller fraction of their ambient power to wake effects. This trend exacerbates the spatial variation in ambient wind power density.

However, there remains a broad distribution of wake-induced losses even amongst locations with similar ambient potentials. The tropics are particularly badly impacted by wake effects, with the mid-latitudes relatively less impacted. That is, for a tropical and mid-latitude location with the same ambient wind potential, the mid-latitude location is likely to suffer from smaller wake-induced losses than the tropical location. This can be explained by the specific wind speed distributions found in each region. The tropics tend to exhibit narrower wind speed distributions than mid-latitude regions, and thus a greater proportion of the wind speed distribution falls between the cut-in and rated speeds of the wind turbine, where power outputs are most sensitive to wake effects. A simple analytic approach derived from one-dimensional momentum theory, based on combining the wind speed distribution with the turbine thrust and power curves, explains 96\% of the variance in simulated wake-induced losses. It follows that wake effects will be sensitive to the choice of turbine model, and therefore that location-dependent selection of turbine models may help to mitigate wake-induced losses.

In assessing wind resources, either for early-stage wind farm planning or to derive inputs into power or energy system models, our results imply that it is not sufficient to assume spatially uniform wake-induced losses. The results also serve to highlight the natural advantages of northern Europe for offshore wind; this region has high ambient potential with relatively low wake-induced losses. In contrast, a number of emerging offshore wind markets are likely to be impacted by these enhanced (relative to mid-latitudes) wake-induced losses, including Brazil, Colombia, Australia and India.

This study also explored the impact of wind direction distribution on the spatial variation in wind power potential. A narrower spread of wind directions presents greater opportunity for the optimisation of array layout designs. The narrowest wind direction distributions tend to be found in the tropics where the wind climate is characterised by the trade winds. Although the narrower wind speed distributions found in these regions lead to stronger wake effects, the narrower wind direction distribution presents greater opportunity for mitigation of these losses via array layout optimisation. Although layout optimisation can only recover a fraction of the potential power lost to wakes, this result reveals the greater importance attached to layout optimisation in the tropics, compared with the mid-latitudes.

Finally, this study assessed future changes in wind power potential and wake-induced losses, due to climate change. Based on the SSP585 scenario for the years 2081--2100, compared with a `historical' period of 1995--2014, much of the globe shows a small but significant decrease in ambient wind potential. This includes Europe and China, the two regions with the greatest current offshore wind capacity. The main exceptions to this are the polar regions and in particular the Arctic, where wind potential is projected to increase. As for the current climate, there is a correlation between changes in ambient wind potential, and percentage power lost to wakes. This means that regions where wind resources decrease due to climate change are also likely to suffer from greater percentage wake losses, exacerbating the climate change effect. Therefore, in assessing climate change impacts on wind potential, it is not sufficient to assume that wake-induced losses will remain the same under a changing climate.

The results of this study have the potential to inform the transfer of knowledge from regions such as Europe, where offshore wind is relatively mature, to emerging markets such as in South America and Southeast Asia. The results highlight the different challenges likely to be experienced in efficiently harnessing wind resources in different regions, as well as the significance of layout optimisation, and potential climate change impacts.

There are a number of avenues for further work. Firstly, this study considered only a single wind farm configuration, and in particular a single turbine model. As discussed above, this has an impact on the simulated wake-induced losses, and warrants further investigation. In particular, by repeating parts of this study for a variety of current or future turbine designs, opportunities to adapt turbine designs to particular wind climates could be explored. Secondly, the layout optimisation in this work has considered only a fixed number of turbines, and optimised the total power generated by the farm. Alternatively, optimising the levelised cost of energy (LCOE) with respect to a variable number of turbines might lead to further mitigation of wake impacts, for example by reducing the installed capacity density in regions more strongly affected by wakes. An LCOE-based optimisation would require a well-validated, globally applicable cost model, which may be a barrier to a global study, but region-specific studies may be feasible. Further cost considerations could also include the impact of increased turbulence due to wake effects on turbine maintenance costs and operating lifetime. Finally, the focus of this study is on intra-farm wake effects. Recent literature has highlighted the significance of inter-farm wake effects, which would add a further spatial (and directional) component to maps of wake-induced losses, as well as additional uncertainty with respect to unknown future build-out. A study of inter-farm wake effects and their uncertainty would require the generation of (an ensemble of) plausible wind farm build-out trajectories. The resulting uncertainties would also need to be considered alongside uncertain climate change.

\section*{Acknowledgements}

We acknowledge HPC resources and support from the Imperial College Research Computing Service (\url{http://doi.org/10.14469/hpc/2232}).

\end{document}